\documentclass[prd,aps,showpacs,epsf,floats,onecolumn]{revtex4}%
\usepackage{amssymb}
\usepackage{amsfonts}
\usepackage{amsmath}
\usepackage{graphicx}%
\providecommand{\U}[1]{\protect\rule{.1in}{.1in}}
\begin{document}
\title{\textbf{Curvature of Quantum Evolutions for Qubits in Time-Dependent Magnetic
Fields}}
\author{\textbf{Carlo Cafaro}$^{1,2}$, \textbf{Leonardo Rossetti}$^{3,1}$,\textbf{
Paul M.\ Alsing}$^{1}$}
\affiliation{$^{1}$University at Albany-SUNY, Albany, NY 12222, USA}
\affiliation{$^{2}$SUNY Polytechnic Institute, Utica, NY 13502, USA}
\affiliation{$^{3}$University of Camerino, I-62032 Camerino, Italy}

\begin{abstract}
In the geometry of quantum-mechanical processes, the time-varying curvature
coefficient of a quantum evolution is specified by the magnitude squared of
the covariant derivative of the tangent vector to the state vector. In
particular, the curvature coefficient measures the bending of the quantum
curve traced out by a parallel-transported pure quantum state that evolves in
a unitary fashion under a nonstationary Hamiltonian that specifies the
Schr\"{o}dinger evolution equation.\textbf{ }In this paper,\textbf{ }we
present an exact analytical expression of the curvature of a quantum evolution
for a two-level quantum system immersed in a time-dependent magnetic field.
Specifically, we study the dynamics generated by a two-parameter nonstationary
Hermitian Hamiltonian with unit speed efficiency. The two parameters specify
the constant temporal rates of change of the polar and azimuthal angles used
in the Bloch sphere representation of the evolving pure state. To better grasp
the physical significance of the curvature coefficient, showing that the
quantum curve is nongeodesic since the geodesic efficiency of the quantum
evolution is strictly less than one and tuning the two Hamiltonian parameters,
we compare the temporal behavior of the curvature coefficient with that of the
speed and the acceleration of the evolution of the system in projective
Hilbert space. Furthermore, we compare the temporal profile of the curvature
coefficient with that of the square of the ratio between the parallel and
transverse magnetic field intensities. Finally, we discuss the\textbf{
}challenges in finding exact analytical solutions when extending our geometric
approach to higher-dimensional quantum systems that evolve unitarily under an
arbitrary time-dependent Hermitian Hamiltonian.

\end{abstract}

\maketitle

\section{Introduction}

The dynamical evolution of quantum systems entails two main challenges
\cite{feynman57}. First, focusing on the Schr\"{o}dinger evolution equation
for two-level systems, the mathematics is not always transparent and the
complex coefficients that appear in the wave function do not give directly the
values of real physical observables. Second, even for two-level systems, there
is a limited applicability of analytical techniques for exactly solving the
Schr\"{o}dinger evolution equation. Therefore, numerical estimations are
necessary in many realistic cases. However, it is generally difficult to gain
general insights about the dependence of the system on its tunable parameters
while performing exclusively numerical investigations. In this respect,
geometrical methods can offer useful tools, since they can help with pictorial
visualizations of quantum interactions. These visualizations, in turn, can
lead to the possibility of gaining deeper insights into the physics of the
problem at hand. Furthermore, geometry offers the chance of handling
relatively simple concepts formulated in terms of ordinary real scalar and
vectorial quantities with a transparent interpretation. Therefore, in
principle, the possibility of exploiting the power of simple geometrical
quantities to characterize aspects of complicated quantum-mechanical phenomena
can be of great theoretical and practical relevance.

Relying on the very clever conversion of the (complex) two-level
Schr\"{o}dinger equation to (real) vectorial three-dimensional spin precession
equation by Feynman and collaborators in Ref. \cite{feynman57}, there have
been fascinating studies on the consequences of mapping quantum-mechanical
two-level systems to a space curve \cite{dandoloff92,carmel00,dandoloff04}. In
this context, making use of suitable methods of differential geometry such as
the Frenet-Serret equations, scientists have provided interesting geometric
descriptions of certain aspects of quantum evolutions by means of the concepts
of curvature and torsion of a space curve. In Ref. \cite{dandoloff92}, for
instance, the curvature of the space curve is related to the variance of the
Hamiltonian. Furthermore, a measure of the deviation of the torsion of the
curve from an instantaneous osculating plane is related to the geometric phase
of the quantum evolution corresponding to the two-level system. In Ref.
\cite{carmel00}, the authors express the curvature and the torsion of the
space curve in terms of the Hamiltonian of the quantum system. Then, from the
study of the behavior of these two geometric quantities, an efficient way of
classifying different families of quantum systems and processes (including,
for instance, two-level atoms, spin-$1/2$ particles, Rosen-Zener systems, and
electronic transitions) was suggested. Finally, using the above-mentioned
Frenet-Serret frame of a space curve, the authors provide in Ref.
\cite{dandoloff04} classical analogues of the Schr\"{o}dinger and Heisenberg
pictures in quantum mechanics. In addition to its clear foundational
relevance, this line of investigation has recently found space for practical
importance in the context of optimal driving of quantum systems with minimal
energy consumption \cite{gentile20}. Currently, the impact that the concepts
of geometric curvature and torsion can have on the future of quantum
technology for superconducting spintronics devices is being seriously
investigated \cite{salamone23}.

In differential geometry, it is established that the classical Frenet-Serret
apparatus of a space curve in three-dimensional Euclidean space specifies the
local geometry of curves \cite{parker77}. In particular, it characterizes
relevant geometric invariants, including the curvature and the torsion of a
curve. Observe that the notion of curvature of a space curve in
three-dimensional Euclidean space as introduced in classical differential
geometry (by the Frenet-Serret curvature coefficient $\kappa_{\mathrm{FS}}$)
is well-defined in a flat space where no gravity is present. In classical
Newtonian mechanics, physicists can describe geometric aspects of classical
trajectories of a particle by means of curvature and torsion coefficients
presented within the Frenet-Serret apparatus. As a side remark, we stress that
the first scientist to consider the geometrization of Newtonian gravity was
Cartan \cite{cartan}\textbf{.} In Ref. \cite{consa18}, for instance, curvature
and torsion coefficients of a space curve were employed to study the geometry
of the cylindrical helix motion of a spin-$1/2$ charged particle in a
homogeneous external magnetic field. Since the curvature of a space curve can
be expressed in terms of velocity and acceleration \cite{parker77}, it is
reasonable to expect that the temporal rate of change of the curvature is
somewhat related to the time-derivative of the acceleration, also known as the
jerk. For an interesting discussion on the jerk of space curves corresponding
to electrons moving in a constant magnetic field in three-dimensional
Euclidean space equipped with moving Frenet-Serret frames, we refer to Refs.
\cite{ozen19,ozen20}.

In geometric quantum mechanics, while analyzing the problem of parameter
estimation, the concept of curvature of a quantum Schr\"{o}dinger trajectory
was introduced in Ref. \cite{brody96} as a generalization of the notion of
curvature of a classical exponential family of distributions of use in
statistical mechanics. In Ref. \cite{brody96}, the curvature of a curve can be
characterized in terms of a suitably introduced squared acceleration vector of
the curve. Furthermore, the curvature measures the parametric sensitivity that
specifies the parametric estimation problem being studied \cite{brody13}. In
Ref. \cite{laba17}, Laba and Tkachuk defined curvature and torsion
coefficients of quantum evolutions for pure states that evolved subject to a
time-independent Hamiltonian evolution. Focusing on single-qubit quantum
states, they showed that the curvature is a measure of the deviation of the
dynamically evolving state vector from the geodesic line on the Bloch sphere
in Ref. \cite{laba17}. In addition, they concluded that the torsion
coefficient \ is a measure of the deviation of the dynamically evolving state
vector from a two-dimensional subspace characterized by the instantaneous
plane of evolution. It is important pointing out that the concepts of
Frenet-Serret curvature, geodesic curvature, and curvature coefficient of a
quantum evolution encode different information about a curve. For instance,
while the curvature of a circle on a plane is an intrinsic property of the
circle, the curvature of a circle on a spherical surface is an extrinsic
property of the circle. For simplicity, take a circle of radius\textbf{ }%
$R$\textbf{. }When this circle is regarded as a great circle on a sphere, it
has geodesic curvature equal to zero.\ Clearly, if the circle lies on the
sphere but is not a great sphere, its geodesic curvature differs from zero.
Moreover, the Frenet-Serret curvature of a circle at the equatorial plane of a
sphere in\textbf{ }$%
\mathbb{R}
^{3}$\textbf{ }is clearly nonzero (unlike what happens when you view the
circle as the equatorial trajectory on a Bloch sphere with the curvature of
the quantum evolution equal to zero). For a discussion on the link among
Frenet-Serret curvature, geodesic curvature, and curvature coefficient of a
quantum evolution, we refer to Ref. \cite{alsing24A}. 

Using the concept of Frenet-Serret apparatus and, in part, influenced by the
investigation by Laba and Tkachuk in Ref. \cite{laba17}, we brought forward in
Refs. \cite{alsing24A, alsing24B,alsing24C} an alternative geometric approach
to specify the bending and the twisting of quantum curves traced out by
dynamically evolving state vectors in a quantum setting. Specifically, we
proposed a quantum version of the Frenet-Serret apparatus for a quantum
trajectory in projective Hilbert space traced out by a parallel-transported
pure quantum state that evolves unitarily under a stationary (or,
alternatively, nonstationary) Hamiltonian specifying the Schr\"{o}dinger
equation. For nonstationary Hamiltonian evolutions \cite{alsing24B}, we found
that the time-varying setting exhibited a richer structure from a statistical
standpoint compared to the stationary setting \cite{alsing24A}. For instance,
unlike the time-independent configuration, we found that the notion of
generalized variance enters nontrivially in the definition of the torsion of a
curve traced out by a quantum state evolving under a nonstationary
Hamiltonian. To physically illustrate the significance of our construct, we
applied it to an exactly soluble time-dependent two-state Rabi problem
specified by a sinusoidal oscillating time-dependent potential. In this
context, we showed that the analytical expressions for the curvature and
torsion coefficients can be (in principle) completely described by only two
real three-dimensional vectors, the Bloch vector that specifies the quantum
system and the externally applied time-varying magnetic field. Although we
demonstrated that the torsion is identically zero for an arbitrary
time-dependent single-qubit Hamiltonian evolution, we only studied in a
numerical fashion the temporal behavior of the curvature coefficient in
different dynamical scenarios, including off-resonance and on-resonance
regimes and, in addition, strong and weak driving configurations.

In this paper, the main goal is that of achieving a deeper understanding of
the physical significance of the curvature coefficient of a quantum evolution
for a two-level quantum system immersed in a time-dependent magnetic field by
providing (for the first time in the literature, to our knowledge) an exact
analytical analysis. Our objective is to handle a spectrum of inquiries,
including but not limited to:

\begin{enumerate}
\item[{[i]}] When transitioning from stationary to nonstationary Hamiltonian
settings, which type of magnetic field configurations yield a vanishing
curvature coefficient?

\item[{[ii]}] Are energy resources necessarily wasted while bending quantum evolutions?

\item[{[iii]}] How is the temporal behavior of the curvature coefficient of a
quantum path related to the temporal profile of its speed and its acceleration
in projective Hilbert space?
\end{enumerate}

Addressing points [i], [ii], and [iii] is relevant for several reasons in the
field of quantum information and computation. For instance, the proper
quantitative understanding of these points can help developing suitable
quantum driving schemes capable of transferring a source state to a target
state in minimal time, optimal speed, and with minimal waste of energy resources.

Before presenting the layout of this paper, we stress that the notions of
\textquotedblleft time-dependent, time-varying, and
nonstationary\textquotedblright\ magnetic field configurations (or,
alternatively, Hamiltonian evolutions) are used as synonyms in this work.

The rest of the paper is organized as follows. In Section II, we present the
essential formal elements yielding the concept of curvature of a quantum
evolution. In Section III, we characterize the dynamics generated by a
two-parameter time-dependent Hamiltonian with unit speed efficiency. The two
parameters determine the constant temporal rates of change of the polar and
azimuthal angles employed in the Bloch sphere description of the
quantum-mechanical evolving pure state. In Section IV, to deepen our
understanding of the concept of curvature, we study several aspects of the
quantum evolution. Specifically, we show that while the nonstationary
Hamiltonian corresponds to a $100\%$ evolution speed efficiency \cite{uzdin12}%
, it generates nongeodesic paths on the Bloch sphere since the geodesic
efficiency of the quantum evolution is strictly less than one
\cite{anandan90,cafaro20}. Furthermore, while tuning the two Hamiltonian
parameters, we study \ and compare the temporal behavior of the speed and the
acceleration of the evolution in projective Hilbert space with that of the
curvature coefficient. Lastly, we provide a comparison between the temporal
profile of the curvature coefficient with that of the square of the ratio
between the parallel (i.e., along the\textbf{ }$z$\textbf{-}axis) and
transverse (i.e., in the\textbf{ }$xy$\textbf{-}plane) magnetic field
intensities. Our summary of findings along with concluding remarks appear in
Section V. Finally, technical details are placed in Appendix A.

\section{Curvature of a quantum evolution}

In this section, we follow our works in Refs.
\cite{alsing24A,alsing24B,alsing24C} to introduce the basic elements leading
to the concept of the curvature of a quantum evolution.

Consider a nonstationary Hamiltonian evolution described by the
Schr\"{o}dinger's equation $i\hslash\partial_{t}\left\vert \psi\left(
t\right)  \right\rangle =\mathrm{H}\left(  t\right)  \left\vert \psi\left(
t\right)  \right\rangle $, with $\left\vert \psi\left(  t\right)
\right\rangle $ belonging to an arbitrary $N$-dimensional complex Hilbert
space $\mathcal{H}_{N}$. Although this introductory presentation assumes that
the reduced Planck constant $\hslash$ is not one, we shall set $\hslash=1$
throughout our paper in most of our explicit calculations. Whenever felt
necessary, the choice of setting $\hslash=1$ will be explicitly reminded to
the reader.\textbf{ }In general, the normalized state vector $\left\vert
\psi\left(  t\right)  \right\rangle $ is such that $\left\langle \psi\left(
t\right)  \left\vert \dot{\psi}\left(  t\right)  \right.  \right\rangle
=(-i/\hslash)\left\langle \psi\left(  t\right)  \left\vert \mathrm{H}\left(
t\right)  \right\vert \psi\left(  t\right)  \right\rangle \neq0$. Given the
state $\left\vert \psi\left(  t\right)  \right\rangle $, we introduce the
parallel transported unit state vector $\left\vert \Psi\left(  t\right)
\right\rangle \overset{\text{def}}{=}e^{i\beta\left(  t\right)  }\left\vert
\psi\left(  t\right)  \right\rangle $ where the phase $\beta\left(  t\right)
$ is such that $\left\langle \Psi\left(  t\right)  \left\vert \dot{\Psi
}\left(  t\right)  \right.  \right\rangle =0$. Observe that $i\hslash
\left\vert \dot{\Psi}\left(  t\right)  \right\rangle =\left[  \mathrm{H}%
\left(  t\right)  -\hslash\dot{\beta}\left(  t\right)  \right]  \left\vert
\Psi\left(  t\right)  \right\rangle $. Then, the relation $\left\langle
\Psi\left(  t\right)  \left\vert \dot{\Psi}\left(  t\right)  \right.
\right\rangle =0$ is equivalent to putting $\beta\left(  t\right)  $ equal to%
\begin{equation}
\beta\left(  t\right)  \overset{\text{def}}{=}\frac{1}{\hslash}\int_{0}%
^{t}\left\langle \psi\left(  t^{\prime}\right)  \left\vert \mathrm{H}\left(
t^{\prime}\right)  \right\vert \psi\left(  t^{\prime}\right)  \right\rangle
dt^{\prime}\text{.}%
\end{equation}
For this reason, $\left\vert \Psi\left(  t\right)  \right\rangle $ reduces to%
\begin{equation}
\left\vert \Psi\left(  t\right)  \right\rangle =e^{(i/\hslash)\int_{0}%
^{t}\left\langle \psi\left(  t^{\prime}\right)  \left\vert \mathrm{H}\left(
t^{\prime}\right)  \right\vert \psi\left(  t^{\prime}\right)  \right\rangle
dt^{\prime}}\left\vert \psi\left(  t\right)  \right\rangle \text{,}%
\label{reason}%
\end{equation}
and satisfies the evolution equation $i\hslash\left\vert \dot{\Psi}\left(
t\right)  \right\rangle =\Delta\mathrm{H}\left(  t\right)  \left\vert
\Psi\left(  t\right)  \right\rangle $ where $\Delta\mathrm{H}\left(  t\right)
\overset{\text{def}}{=}\mathrm{H}\left(  t\right)  -\left\langle
\mathrm{H}\left(  t\right)  \right\rangle $. As pointed out in Ref.
\cite{alsing24B}, the speed $v(t)$ of quantum evolution is not constant when
the Hamiltonian is nonstationary. In particular, $v(t)$ is such that
$v^{2}\left(  t\right)  =\left\langle \dot{\Psi}\left(  t\right)  \left\vert
\dot{\Psi}\left(  t\right)  \right.  \right\rangle =\left\langle \left(
\Delta\mathrm{H}\left(  t\right)  \right)  ^{2}\right\rangle /\hslash^{2}$. We
introduce for convenience the arc length $s=s\left(  t\right)  $ specified by
means of $v\left(  t\right)  $ as%
\begin{equation}
s\left(  t\right)  \overset{\text{def}}{=}\int_{0}^{t}v(t^{\prime})dt^{\prime
}\text{,}\label{s-equation}%
\end{equation}
with $ds=v(t)dt$, that is, $\partial_{t}=v(t)\partial_{s}$. Clearly,
$\partial_{t}\overset{\text{def}}{=}\partial/\partial t$ and $\partial
_{s}\overset{\text{def}}{=}\partial/\partial s$. Finally, presenting the
adimensional operator%
\begin{equation}
\Delta h\left(  t\right)  \overset{\text{def}}{=}\frac{\Delta\mathrm{H}\left(
t\right)  }{\hslash v(t)}=\frac{\Delta\mathrm{H}\left(  t\right)  }%
{\sqrt{\left\langle \left(  \Delta\mathrm{H}\left(  t\right)  \right)
^{2}\right\rangle }}\text{,}%
\end{equation}
the normalized tangent vector $\left\vert T\left(  s\right)  \right\rangle
\overset{\text{def}}{=}\partial_{s}\left\vert \Psi\left(  s\right)
\right\rangle =\left\vert \Psi^{\prime}\left(  s\right)  \right\rangle $
becomes $\left\vert T\left(  s\right)  \right\rangle =-i\Delta h\left(
s\right)  \left\vert \Psi\left(  s\right)  \right\rangle $. We remark that
$\left\langle T\left(  s\right)  \left\vert T\left(  s\right)  \right.
\right\rangle =1$ by construction and, in addition, $\partial_{s}\left\langle
\Delta h(s)\right\rangle =\left\langle \Delta h^{\prime}(s)\right\rangle $.
Indeed, this latter relation can be extended to an arbitrary power of
differentiation. To the second power, for example, we have $\partial_{s}%
^{2}\left\langle \Delta h(s)\right\rangle =\left\langle \Delta h^{\prime
\prime}(s)\right\rangle $. We can construct $\left\vert T^{\prime}\left(
s\right)  \right\rangle \overset{\text{def}}{=}\partial_{s}\left\vert T\left(
s\right)  \right\rangle $ from the tangent vector $\left\vert T\left(
s\right)  \right\rangle =-i\Delta h\left(  s\right)  \left\vert \Psi\left(
s\right)  \right\rangle $. As a matter of fact, after some algebra, we obtain
$\left\vert T^{\prime}\left(  s\right)  \right\rangle =-i\Delta h(s)\left\vert
\Psi^{\prime}\left(  s\right)  \right\rangle -i\Delta h^{\prime}(s)\left\vert
\Psi\left(  s\right)  \right\rangle $ where
\begin{equation}
\left\langle T^{\prime}\left(  s\right)  \left\vert T^{\prime}\left(
s\right)  \right.  \right\rangle =\left\langle \left(  \Delta h^{\prime
}(s)\right)  ^{2}\right\rangle +\left\langle \left(  \Delta h(s)\right)
^{4}\right\rangle -2i\operatorname{Re}\left[  \left\langle \Delta h^{\prime
}(s)\left(  \Delta h(s)\right)  ^{2}\right\rangle \right]  \neq1\text{,}%
\end{equation}
in general. We are now prepared to present the curvature coefficient for
quantum evolutions generated by nonstationary Hamiltonians. Indeed, having
introduced the vectors $\left\vert \Psi\left(  s\right)  \right\rangle $,
$\left\vert T\left(  s\right)  \right\rangle $, and $\left\vert T^{\prime
}\left(  s\right)  \right\rangle $, we can finally establish the curvature
coefficient $\kappa_{\mathrm{AC}}^{2}\left(  s\right)  \overset{\text{def}}%
{=}\left\langle \tilde{N}_{\ast}\left(  s\right)  \left\vert \tilde{N}_{\ast
}\left(  s\right)  \right.  \right\rangle $ with $\left\vert \tilde{N}_{\ast
}\left(  s\right)  \right\rangle \overset{\text{def}}{=}\mathrm{P}^{\left(
\Psi\right)  }\left\vert T^{\prime}\left(  s\right)  \right\rangle $, where
the projection operator $\mathrm{P}^{\left(  \Psi\right)  }$ onto states
orthogonal to $\left\vert \Psi\left(  s\right)  \right\rangle $ is given
by\textbf{ }$\mathrm{P}^{\left(  \Psi\right)  }\overset{\text{def}}%
{=}\mathrm{I}-\left\vert \Psi\left(  s\right)  \right\rangle \left\langle
\Psi\left(  s\right)  \right\vert $, and \textquotedblleft$\mathrm{I}%
$\textquotedblright\ denotes the identity operator in $\mathcal{H}_{N}$. As
mentioned in Refs. \cite{alsing24A,alsing24B}, the subscript \textquotedblleft%
\textrm{AC}\textquotedblright\textrm{\ }means Alsing and Cafaro. \ Observe
that the curvature coefficient $\kappa_{\mathrm{AC}}^{2}\left(  s\right)  $
can be recast as%
\begin{equation}
\kappa_{\mathrm{AC}}^{2}\left(  s\right)  \overset{\text{def}}{=}\left\Vert
\mathrm{D}\left\vert T(s)\right\rangle \right\Vert ^{2}=\left\Vert
\mathrm{D}^{2}\left\vert \Psi\left(  s\right)  \right\rangle \right\Vert
^{2}\text{,}\label{peggio}%
\end{equation}
where $\mathrm{D}\overset{\text{def}}{=}\mathrm{P}^{\left(  \Psi\right)
}d/ds=\left(  \mathrm{I}-\left\vert \Psi\right\rangle \left\langle
\Psi\right\vert \right)  d/ds$ with $\mathrm{D}\left\vert T(s)\right\rangle
\overset{\text{def}}{=}\mathrm{P}^{\left(  \Psi\right)  }\left\vert T^{\prime
}(s)\right\rangle =$ $\left\vert \tilde{N}_{\ast}\left(  s\right)
\right\rangle $ representing the covariant derivative
\cite{carlocqg23,samuel88,paulPRA23}. In this paper, we choose to use
$\left\vert \tilde{N}_{\ast}\left(  s\right)  \right\rangle $ to
express\textbf{ }$\kappa_{\mathrm{AC}}^{2}$ when we wish to avoid cumbersome
expressions needed to define the covariant derivative $\mathrm{D}$ and the
projection operator $\mathrm{P}^{\left(  \Psi\right)  }$. From Eq.
(\ref{peggio}), we note that the curvature coefficient $\kappa_{\mathrm{AC}%
}^{2}\left(  s\right)  $ is equal to the magnitude squared of the second
covariant derivative of the state vector $\left\vert \Psi\left(  s\right)
\right\rangle $ that delineates the quantum Schr\"{o}dinger trajectory in
projective Hilbert space. For clarity, we emphasize that $\left\vert \tilde
{N}_{\ast}\left(  s\right)  \right\rangle $ is a vector that is neither
orthogonal to the vector $\left\vert T\left(  s\right)  \right\rangle $ nor
normalized to one. Instead, despite not being properly normalized, $\left\vert
\tilde{N}\left(  s\right)  \right\rangle \overset{\text{def}}{=}%
\mathrm{P}^{\left(  T\right)  }\mathrm{P}^{\left(  \Psi\right)  }\left\vert
T^{\prime}\left(  s\right)  \right\rangle $ is orthogonal to $\left\vert
T\left(  s\right)  \right\rangle $. Finally, $\left\vert N\left(  s\right)
\right\rangle \overset{\text{def}}{=}$ $\left\vert \tilde{N}\left(  s\right)
\right\rangle /\sqrt{\left\langle \tilde{N}\left(  s\right)  \left\vert
\tilde{N}\left(  s\right)  \right.  \right\rangle }$ describes a normalized
vector that is also orthogonal to $\left\vert T\left(  s\right)  \right\rangle
$. In conclusion, $\left\{  \left\vert \Psi\left(  s\right)  \right\rangle
\text{, }\left\vert T\left(  s\right)  \right\rangle \text{, }\left\vert
N\left(  s\right)  \right\rangle \right\}  $ represents the set of three
orthonormal vectors required to characterize the curvature of a quantum
evolution. Observe that we focus our attention on the three-dimensional
complex subspace spanned by $\left\{  \left\vert \Psi\left(  s\right)
\right\rangle \text{, }\left\vert T\left(  s\right)  \right\rangle \text{,
}\left\vert N\left(  s\right)  \right\rangle \right\}  $, although
$\mathcal{H}_{N}$ can have arbitrary dimension $N$ as a complex space.
Nevertheless, our choice is consistent with the classical geometric
perspective where the curvature and torsion coefficients can be viewed as the
lowest and second-lowest, respectively, members of a family of generalized
curvature functions \cite{alvarez19}. In particular, for curves in
higher-dimensional spaces, this well grounded geometric viewpoint requires a
set of $m$ orthonormal vectors in order to create $\left(  m-1\right)
$-generalized curvature functions \cite{alvarez19}.

The explicit calculation of the time-dependent curvature coefficient
$\kappa_{\mathrm{AC}}^{2}\left(  s\right)  $ in Eqs. (\ref{peggio}) based on
the projection operators formalism is usually problematic. The challenge is
due to the fact that, similarly to the classical case of space curves in $%
\mathbb{R}
^{3}$ \cite{parker77}, there are essentially two drawbacks during the
reparametrization of a quantum curve by its arc length $s$. First, we may not
be able to calculate in closed form $s\left(  t\right)  $ in Eq.
(\ref{s-equation}). Second, even if we are able to get $s=s\left(  t\right)
$, we may be unable to invert this relation and, thus, arrive at $t=t\left(
s\right)  $ needed to express $\left\vert \Psi\left(  s\right)  \right\rangle
\overset{\text{def}}{=}\left\vert \Psi\left(  t(s)\right)  \right\rangle $. To
bypass these hurdles, we can recast $\kappa_{\mathrm{AC}}^{2}\left(  s\right)
$ in Eq. (\ref{peggio}) by means of expectation values taken with respect to
the state $\left\vert \Psi\left(  t\right)  \right\rangle $ (or,
alternatively, with respect to $\left\vert \psi\left(  t\right)  \right\rangle
$) that can be calculated with no need of the relation $t=t\left(  s\right)
$\textbf{ }\cite{alsing24A,alsing24B}. For simplicity, we shall make no
explicit reference to the the $s$-dependence of the various operators and
expectation values in the following discussion. For example, $\Delta h\left(
s\right)  $ will be written as $\Delta h$. After some algebraic manipulations,
we obtain%
\begin{equation}
\left\vert \tilde{N}_{\ast}\right\rangle =-\left\{  \left[  \left(  \Delta
h\right)  ^{2}-\left\langle \left(  \Delta h\right)  ^{2}\right\rangle
\right]  +i\left[  \Delta h^{\prime}-\left\langle \Delta h^{\prime
}\right\rangle \right]  \right\}  \left\vert \Psi\right\rangle \text{,}
\label{nas}%
\end{equation}
where $\Delta h^{\prime}=\partial_{s}\left(  \Delta h\right)  =\left[
\partial_{t}\left(  \Delta h\right)  \right]  /v\left(  t\right)  $. To
evaluate $\kappa_{\mathrm{AC}}^{2}\left(  s\right)  \overset{\text{def}}%
{=}\left\langle \tilde{N}_{\ast}\left(  s\right)  \left\vert \tilde{N}_{\ast
}\left(  s\right)  \right.  \right\rangle $, it is useful to bring in the
Hermitian operator $\hat{\alpha}_{1}\overset{\text{def}}{=}\left(  \Delta
h\right)  ^{2}-\left\langle \left(  \Delta h\right)  ^{2}\right\rangle $ and
the anti-Hermitian operator $\hat{\beta}_{1}\overset{\text{def}}{=}i\left[
\Delta h^{\prime}-\left\langle \Delta h^{\prime}\right\rangle \right]  $,
where $\hat{\beta}_{1}^{\dagger}=-\hat{\beta}_{1}$. Then, $\left\vert
\tilde{N}_{\ast}\right\rangle =-\left(  \hat{\alpha}_{1}+\hat{\beta}%
_{1}\right)  \left\vert \Psi\right\rangle $ and $\left\langle \tilde{N}_{\ast
}\left(  s\right)  \left\vert \tilde{N}_{\ast}\left(  s\right)  \right.
\right\rangle $ is equal to $\left\langle \hat{\alpha}_{1}^{2}\right\rangle
-\left\langle \hat{\beta}_{1}^{2}\right\rangle +\left\langle \left[
\hat{\alpha}_{1}\text{, }\hat{\beta}_{1}\right]  \right\rangle $, where
$\left[  \hat{\alpha}_{1}\text{, }\hat{\beta}_{1}\right]  \overset{\text{def}%
}{=}\hat{\alpha}_{1}\hat{\beta}_{1}-\hat{\beta}_{1}\hat{\alpha}_{1}$ denotes
the quantum commutator of $\hat{\alpha}_{1}$ and $\hat{\beta}_{1}$. Observe
that the expectation value $\left\langle \left[  \hat{\alpha}_{1}\text{, }%
\hat{\beta}_{1}\right]  \right\rangle $ is a real number since $\left[
\hat{\alpha}_{1}\text{, }\hat{\beta}_{1}\right]  $ is a Hermitian operator.
This is a consequence of the fact that $\hat{\alpha}_{1}$ and $\hat{\beta}%
_{1}$ are Hermitian and anti-Hermitian operators, respectively. Employing the
definitions of $\hat{\alpha}_{1}$ and $\hat{\beta}_{1}$, we get $\left\langle
\hat{\alpha}_{1}^{2}\right\rangle =\left\langle (\Delta h)^{4}\right\rangle
-\left\langle (\Delta h)^{2}\right\rangle ^{2}$, $\left\langle \hat{\beta}%
_{1}^{2}\right\rangle =-\left[  \left\langle (\Delta h^{\prime})^{2}%
\right\rangle -\left\langle \Delta h^{\prime}\right\rangle ^{2}\right]  $, and
$\left\langle \left[  \hat{\alpha}_{1}\text{, }\hat{\beta}_{1}\right]
\right\rangle =i\left\langle \left[  (\Delta h)^{2}\text{, }\Delta h^{\prime
}\right]  \right\rangle $. Note that, since $\left[  (\Delta h)^{2}\text{,
}\Delta h^{\prime}\right]  $ is a anti-Hermitian operator, $\left\langle
\left[  (\Delta h)^{2}\text{, }\Delta h^{\prime}\right]  \right\rangle $ is
purely imaginary. For completeness, we stress that $\left[  (\Delta
h)^{2}\text{, }\Delta h^{\prime}\right]  $ is not usually a null operator.
Indeed, $\left[  (\Delta h)^{2}\text{, }\Delta h^{\prime}\right]  =\Delta
h\left[  \Delta h\text{, }\Delta h^{\prime}\right]  +\left[  \Delta h\text{,
}\Delta h^{\prime}\right]  \Delta h$ where $\left[  \Delta h\text{, }\Delta
h^{\prime}\right]  =\left[  \mathrm{H}\text{, }\mathrm{H}^{\prime}\right]  $.
Then, limiting our attention to nonstationary qubit Hamiltonians of the form
\textrm{H}$\left(  s\right)  \overset{\text{def}}{=}\mathbf{h}\left(
s\right)  \cdot\mathbf{\boldsymbol{\sigma}}$, the commutator $\left[
\mathrm{H}\text{, }\mathrm{H}^{\prime}\right]  =2i(\mathbf{h}\times
\mathbf{h}^{\prime})\cdot\mathbf{\boldsymbol{\sigma}}$ can be nonzero since
the vectors $\mathbf{h}$ and $\mathbf{h}^{\prime}$ are not generally
collinear. Clearly, $\mathbf{\boldsymbol{\sigma}}$ denotes the vector operator
whose components are specified by the usual Pauli operators $\sigma_{x}$,
$\sigma_{y}$, and $\sigma_{z}$. Finally, a computationally convenient
expression for the curvature coefficient $\kappa_{\mathrm{AC}}^{2}\left(
s\right)  $ in Eq. (\ref{peggio}) in an arbitrary nonstationary setting
reduces to%
\begin{equation}
\kappa_{\mathrm{AC}}^{2}\left(  s\right)  =\left\langle (\Delta h)^{4}%
\right\rangle -\left\langle (\Delta h)^{2}\right\rangle ^{2}+\left[
\left\langle (\Delta h^{\prime})^{2}\right\rangle -\left\langle \Delta
h^{\prime}\right\rangle ^{2}\right]  +i\left\langle \left[  (\Delta
h)^{2}\text{, }\Delta h^{\prime}\right]  \right\rangle \text{.}
\label{curvatime}%
\end{equation}
From Eq. (\ref{curvatime}), we note that when the Hamiltonian \textrm{H} is
constant, $\Delta h^{\prime}$ becomes the null operator and we recover the
stationary limit $\left\langle (\Delta h)^{4}\right\rangle -\left\langle
(\Delta h)^{2}\right\rangle ^{2}$ of $\kappa_{\mathrm{AC}}^{2}\left(
s\right)  $ \cite{alsing24A}.

The expression of $\kappa_{\mathrm{AC}}^{2}\left(  s\right)  $ in Eq.
(\ref{curvatime}) is obtained via an approach that relies upon the calculation
of expectation values which, in turn, require the knowledge of \ the state
vector $\left\vert \psi\left(  t\right)  \right\rangle $ governed by the
time-dependent Schr\"{o}dinger's evolution equation. As argued in Ref.
\cite{alsing24B}, such expectation-values approach offers an insightful
statistical interpretation for $\kappa_{\mathrm{AC}}^{2}\left(  s\right)  $.
At the same time, however, it is devoid of a clear geometrical interpretation.
Motivated by this deficiency and focusing on nonstationary Hamiltonians and
two-level quantum systems, it is possible to find a closed-form expression for
the curvature coefficient for a curve mapped out by a single-qubit quantum
state that evolves under an arbitrary nonstationary Hamiltonian. The curvature
coefficient $\kappa_{\mathrm{AC}}^{2}$ can be expressed completely in terms of
exclusively two real three-dimensional vectors with a transparent geometric
significance. Specifically, the two vectors are the Bloch vector
$\mathbf{a}\left(  t\right)  $ and the magnetic field vector $\mathbf{h}%
\left(  t\right)  $. While the former vector emerges from the density operator
$\rho\left(  t\right)  =$ $\left\vert \psi\left(  t\right)  \right\rangle
\left\langle \psi\left(  t\right)  \right\vert \overset{\text{def}}{=}\left[
\mathrm{I}+\mathbf{a}\left(  t\right)  \cdot\mathbf{\boldsymbol{\sigma}%
}\right]  /2$, the latter specifies the nonstationary Hamiltonian
\textrm{H}$\left(  t\right)  \overset{\text{def}}{=}\mathbf{h}\left(
t\right)  \cdot\mathbf{\boldsymbol{\sigma}}$. Following the detailed analysis
in Ref. \cite{alsing24B}, we get%
\begin{equation}
\kappa_{\mathrm{AC}}^{2}\left(  \mathbf{a}\text{, }\mathbf{h}\right)
=4\frac{\left(  \mathbf{a\cdot h}\right)  ^{2}}{\mathbf{h}^{2}-\left(
\mathbf{a\cdot h}\right)  ^{2}}+\frac{\left[  \mathbf{h}^{2}\mathbf{\dot{h}%
}^{2}-\left(  \mathbf{h\cdot\dot{h}}\right)  ^{2}\right]  -\left[  \left(
\mathbf{a\cdot\dot{h}}\right)  \mathbf{h-}\left(  \mathbf{a\cdot h}\right)
\mathbf{\dot{h}}\right]  ^{2}}{\left[  \mathbf{h}^{2}-\left(  \mathbf{a\cdot
h}\right)  ^{2}\right]  ^{3}}+4\frac{\left(  \mathbf{a\cdot h}\right)  \left[
\mathbf{a\cdot}\left(  \mathbf{h\times\dot{h}}\right)  \right]  }{\left[
\mathbf{h}^{2}-\left(  \mathbf{a\cdot h}\right)  ^{2}\right]  ^{2}}\text{.}
\label{XXX}%
\end{equation}
The expression of $\kappa_{\mathrm{AC}}^{2}$ in Eq. (\ref{XXX}) is very useful
from a computational standpoint for qubit systems and, at the same time,
offers a clear geometric interpretation of the curvature of a quantum
evolution in terms of the (normalized unitless) Bloch vector $\mathbf{a}$ and
the\ (generally unnormalized, with $\left[  \mathbf{h}\right]  _{\mathrm{MKSA}%
}=$\textrm{joules}$=\sec.^{-1}$ when setting $\hslash=1$) magnetic field
vector $\mathbf{h}$.

Having introduced the curvature coefficient $\kappa_{\mathrm{AC}}^{2}\left(
\mathbf{a}\text{, }\mathbf{h}\right)  $ in Eq. (\ref{XXX}), we can finally
introduce the time-dependent Hamiltonian $\mathrm{H}\left(  t\right)
=\mathbf{h}\left(  t\right)  \cdot\mathbf{\boldsymbol{\sigma}}$ in what follows.

\section{The nonstationary Hamiltonian}

In this section, we specify the dynamics governed by a two-parameter
nonstationary Hamiltonian with unit speed efficiency.

\subsection{Preliminaries}

In Ref. \cite{uzdin12}, Uzdin and collaborators introduced suitable families
of time-dependent Hamiltonians capable\textbf{ }of producing predefined
trajectories, not necessarily geodesic paths, with minimal waste of energetic
resources. In particular, the condition of minimal waste of energetic
resources is achieved when no energy is wasted on parts of the Hamiltonian
that do not actively drive the system. In other words, all the available
energy quantified in terms of the spectral norm of the Hamiltonian $\left\Vert
\mathrm{H}\right\Vert _{\mathrm{SP}}$ is converted into the speed of evolution
of the system quantified by the energy uncertainty $\Delta E$, $v_{\mathrm{H}%
}(t)\overset{\text{def}}{=}(\gamma/\hslash)\Delta E\left(  t\right)  $.
Observe that $\gamma$ is an arbitrary positive constant whose chosen value
depends on how the Fubini-Study metric is defined. For instance, while
$\gamma=1$ in Ref. \cite{uzdin12}, $\gamma=2$ in Ref. \cite{laba17}. More
specifically, the so-called Uzdin's speed efficiency $\eta_{\mathrm{SE}}%
=\eta_{\mathrm{SE}}\left(  t\right)  $ is defined as \cite{uzdin12}%
\begin{equation}
\eta_{\mathrm{SE}}\overset{\text{def}}{=}\frac{\Delta\mathrm{H}_{\rho}%
}{\left\Vert \mathrm{H}\right\Vert _{\mathrm{SP}}}=\frac{\sqrt{\mathrm{tr}%
\left(  \rho\mathrm{H}^{2}\right)  -\left[  \mathrm{tr}\left(  \rho
\mathrm{H}\right)  \right]  ^{2}}}{\max\left[  \sqrt{\mathrm{eig}\left(
\mathrm{H}^{\dagger}\mathrm{H}\right)  }\right]  }\text{,} \label{se1}%
\end{equation}
where $\left\Vert \mathrm{H}\right\Vert _{\mathrm{SP}}\overset{\text{def}}%
{=}\max\left[  \sqrt{\mathrm{eig}\left(  \mathrm{H}^{\dagger}\mathrm{H}%
\right)  }\right]  $ is the so-called spectral norm of the Hamiltonian
operator \textrm{H }and $\rho$ is the density operator that describes the
system. The spectral norm measures the size of bounded linear operators and is
defined as the square root of the maximum eigenvalue of the operator
$\mathrm{H}^{\dagger}\mathrm{H}$, with $\mathrm{H}^{\dagger}$ being the
Hermitian adjoint of $\mathrm{H}$. In what follows, we focus on qubit systems.
Then, after recasting the generally time-dependent Hamiltonian in Eq.
(\ref{se1}) as $\mathrm{H}\left(  t\right)  =h_{0}\left(  t\right)
\mathrm{I}+\mathbf{h}\left(  t\right)  \cdot\mathbf{\boldsymbol{\sigma}}$, we
note that $\eta_{\mathrm{SE}}$ in Eq. (\ref{se1}) can be formally recast as%
\begin{equation}
\eta_{\mathrm{SE}}=\frac{\sqrt{\mathbf{h\cdot h}-(\mathbf{a}\cdot
\mathbf{h})^{2}}}{\left\vert h_{0}\right\vert +\sqrt{\mathbf{h\cdot h}}%
}\text{,} \label{se2}%
\end{equation}
with $\mathbf{a=a}\left(  t\right)  $ being the instantaneous unit Bloch
vector that specifies the qubit state of the system, and
\begin{equation}
\mathrm{eig}\left(  \mathrm{H}^{\dagger}\mathrm{H}\right)  =\left\{
\lambda_{\mathrm{H}^{\dagger}\mathrm{H}}^{\left(  +\right)  }\overset
{\text{def}}{=}(h_{0}+\sqrt{\mathbf{h\cdot h}})^{2}\text{, }\lambda
_{\mathrm{H}^{\dagger}\mathrm{H}}^{\left(  -\right)  }\overset{\text{def}}%
{=}(h_{0}-\sqrt{\mathbf{h\cdot h}})^{2}\right\}  \text{.}%
\end{equation}
For completeness, observe that $\eta_{\mathrm{SE}}=\eta_{\mathrm{SE}}\left(
t\right)  \in\left[  0\text{, }1\right]  $ is a local measure of efficiency
for the quantum system. Furthermore, given that the eigenvalues of
$\mathrm{H}\left(  t\right)  =h_{0}\left(  t\right)  \mathrm{I}+\mathbf{h}%
\left(  t\right)  \cdot\mathbf{\boldsymbol{\sigma}}$ are given by $E_{\pm
}\overset{\text{def}}{=}h_{0}\pm\sqrt{\mathbf{h\cdot h}}$, we have that
$h_{0}=(E_{+}+E_{-})/2$ is the average of the two energy levels. The quantity
$\sqrt{\mathbf{h\cdot h}}=(E_{+}-E_{-})/2$, instead, is proportional to the
energy splitting between the two energy levels $E_{\pm}$. Finally, for a
traceless nonstationary Hamiltonian $\mathrm{H}\left(  t\right)
=\mathbf{h}\left(  t\right)  \cdot\mathbf{\boldsymbol{\sigma}}$, with
$\mathbf{a}(t)\cdot\mathbf{h}\left(  t\right)  =0$ for any instant $t$,
$\eta_{\mathrm{SE}}\left(  t\right)  =1$ and the quantum evolution occurs with
no waste of energetic resources.

What is the most general Hermitian nonstationary qubit Hamiltonian
$\mathrm{H}\left(  t\right)  $ proposed in Ref. \cite{uzdin12}? The
Hamiltonian $\mathrm{H}\left(  t\right)  $ is constructed in such a manner
that it generates the same motion $\pi\left(  \left\vert \psi\left(  t\right)
\right\rangle \right)  $ in the complex projective Hilbert space $%
\mathbb{C}
P^{1}$ (or, equivalently, on the Bloch sphere $S^{2}\cong%
\mathbb{C}
P^{1}$) as $\left\vert \psi\left(  t\right)  \right\rangle $, where the
projection operator $\pi$ is such that $\pi:\mathcal{H}_{2}^{1}\ni\left\vert
\psi\left(  t\right)  \right\rangle \mapsto\pi\left(  \left\vert \psi\left(
t\right)  \right\rangle \right)  \in%
\mathbb{C}
P^{1}$. In general, it can be shown that $\mathrm{H}\left(  t\right)  $ can be
written as \cite{uzdin12}%
\begin{equation}
\mathrm{H}\left(  t\right)  =i\left\vert \partial_{t}m(t)\right\rangle
\left\langle m(t)\right\vert -i\left\vert m(t)\right\rangle \left\langle
\partial_{t}m(t)\right\vert \text{,}\label{optH}%
\end{equation}
where, for simplicity of notation, we can set $\left\vert m(t)\right\rangle
=\left\vert m\right\rangle $ and $\left\vert \partial_{t}m(t)\right\rangle
=\left\vert \partial_{t}m\right\rangle =\partial_{t}\left\vert m\right\rangle
=\left\vert \dot{m}\right\rangle $. The state $\left\vert m\right\rangle $ is
such that $\pi\left(  \left\vert m(t)\right\rangle \right)  =\pi\left(
\left\vert \psi\left(  t\right)  \right\rangle \right)  $, $i\partial
_{t}\left\vert m(t)\right\rangle =\mathrm{H}\left(  t\right)  \left\vert
m(t)\right\rangle $, and $\eta_{\mathrm{SE}}\left(  t\right)  =1$. The
requirement that $\pi\left(  \left\vert m(t)\right\rangle \right)  =\pi\left(
\left\vert \psi\left(  t\right)  \right\rangle \right)  $ implies that
$\left\vert m(t)\right\rangle =c(t)\left\vert \psi\left(  t\right)
\right\rangle $ for some complex function $c(t)$. Imposing that $\left\langle
m\left\vert m\right.  \right\rangle =1$, we get $\left\vert c(t)\right\vert
=1$. Therefore, $c(t)=e^{i\phi\left(  t\right)  }$ for some real function
$\phi\left(  t\right)  $. Then, imposing the parallel transport condition
$\left\langle m\left\vert \dot{m}\right.  \right\rangle =\left\langle \dot
{m}\left\vert m\right.  \right\rangle =0$, the phase $\phi\left(  t\right)  $
becomes equal to $i\int\left\langle \psi\left\vert \dot{\psi}\right.
\right\rangle dt$. Therefore, $\left\vert m(t)\right\rangle =\exp(-\int
_{0}^{t}\left\langle \psi(t^{\prime})\left\vert \partial_{t^{\prime}}%
\psi(t^{\prime})\right.  \right\rangle dt^{\prime})\left\vert \psi\left(
t\right)  \right\rangle $. Observe that $\left\vert m(t)\right\rangle $ is the
same as the parallel transported unit state vector $\left\vert \Psi\left(
t\right)  \right\rangle $ introduced in Eq. (\ref{reason}). However, in what
follows, we keep the original notation employed in Ref. \cite{uzdin12} and use
$\left\vert m(t)\right\rangle $ instead of\textbf{ }$\left\vert \Psi\left(
t\right)  \right\rangle $. Note that $\mathrm{H}\left(  t\right)  $ in Eq.
(\ref{optH}) is traceless by construction since it possesses only off-diagonal
elements in the orthogonal basis $\left\{  \left\vert m\right\rangle \text{,
}\left\vert \partial_{t}m\right\rangle \right\}  $. Furthermore, the condition
$i\partial_{t}\left\vert m(t)\right\rangle =\mathrm{H}\left(  t\right)
\left\vert m(t)\right\rangle $ implies that $\left\vert m(t)\right\rangle $
satisfies the Schr\"{o}dinger evolution equation. Finally, the relation
$\eta_{\mathrm{SE}}\left(  t\right)  =1$ signifies that $\mathrm{H}\left(
t\right)  $ drives $\left\vert m(t)\right\rangle $ with maximal speed with no
waste of energy resources.

Having presented some basic preliminary material on Uzdin's work in Ref.
\cite{uzdin12}, we can now introduce our proposed time-dependent unit speed
efficiency Hamiltonian.\begin{figure}[t]
\centering
\includegraphics[width=0.35\textwidth] {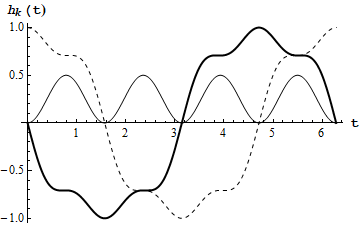}\caption{Visualization of the
temporal behavior of the magnetic field components $\left\{  h_{k}\right\}
_{1\leq k\leq3}$ with $h_{1}=h_{x}$ (thick solid), $h_{2}=h_{y}$ (dashed), and
$h_{3}=h_{z}$ (thin solid). In the plot, we set $\omega_{0}=\nu_{0}=1$. Note
that, in this case, $h_{y}(t)=h_{x}(t-\pi/2)$. Physical units are selected so
that $\hslash=1$.}%
\end{figure}

\subsection{The Hamiltonian}

In what follows, we aim to construct a time-dependent qubit Hamiltonian with
$100\%$ evolution speed that drives the quantum system with maximal speed
efficiency (i.e., $\eta_{\mathrm{SE}}=1$, with $\eta_{\mathrm{SE}}$ in Eq.
(\ref{se1})) along a non-geodesic path (i.e., $0\leq\eta_{\mathrm{GE}}<1$,
with $\eta_{\mathrm{GE}}$ in Eq. (\ref{eta1})) on the Bloch sphere. Therefore,
we expect the curvature $\kappa_{\mathrm{AC}}^{2}(t)$ of the quantum evolution
to be nonvanishing and, in principle, to exhibit a nonconstant temporal behavior.

We begin by considering a unit quantum state given by $\left\vert \psi\left(
t\right)  \right\rangle =\cos\left(  \omega_{0}t\right)  \left\vert
0\right\rangle +e^{i\nu_{0}t}\sin\left(  \omega_{0}t\right)  \left\vert
1\right\rangle $, with $\nu_{0}$ and $\omega_{0}$ being the two essential real
positive parameters that can be tuned when studying the quantum evolution.
Recall that in terms of the polar and azimuthal angles $\theta\left(
t\right)  $ and $\varphi\left(  t\right)  $, respectively, an arbitrary state
on the Bloch sphere can be recast as $\left\vert \psi\left(  t\right)
\right\rangle =\left\vert \psi\left(  \theta\left(  t\right)  \text{, }%
\varphi\left(  t\right)  \right)  \right\rangle =\cos\left[  \theta\left(
t\right)  /2\right]  \left\vert 0\right\rangle +e^{i\varphi\left(  t\right)
}\sin\left[  \theta\left(  t\right)  /2\right]  \left\vert 1\right\rangle $.
Therefore, we have $\omega_{0}t=\theta\left(  t\right)  /2$ and $\nu
_{0}t=\varphi\left(  t\right)  $, that is, $\omega_{0}=\dot{\theta}/2$ and
$\nu_{0}=\dot{\varphi}$. Since $\left\langle \psi\left(  t\right)  \left\vert
\dot{\psi}\left(  t\right)  \right.  \right\rangle =i\nu_{0}\sin^{2}\left(
\omega_{0}t\right)  \neq0$, $\left\vert \psi\left(  t\right)  \right\rangle $
is not parallel transported. From the state $\left\vert \psi\left(  t\right)
\right\rangle $, we consider the state $\left\vert m(t)\right\rangle
\overset{\text{def}}{=}e^{-i\phi\left(  t\right)  }\left\vert \psi\left(
t\right)  \right\rangle $ with the phase $\phi\left(  t\right)  $ such that
$\left\langle m(t)\left\vert \dot{m}(t)\right.  \right\rangle =0$. A simple
calculation implies that $\left\langle m(t)\left\vert \dot{m}(t)\right.
\right\rangle =0$ if and only if $-i\dot{\phi}+\left\langle \psi\left(
t\right)  \left\vert \dot{\psi}\left(  t\right)  \right.  \right\rangle =0$,
that is $\dot{\phi}=-i\left\langle \psi\left(  t\right)  \left\vert \dot{\psi
}\left(  t\right)  \right.  \right\rangle $. Setting $\phi\left(  0\right)
=0$, the temporal behavior of the phase $\phi\left(  t\right)  $ is given by
\begin{equation}
\phi\left(  t\right)  =\frac{\nu_{0}}{4\omega_{0}}\left[  2\omega_{0}%
t-\sin\left(  2\omega_{0}t\right)  \right]  \text{.}\label{s5}%
\end{equation}
Furthermore, using Eq. (\ref{s5}), the parallel transported unit state
$\left\vert m(t)\right\rangle $ becomes%
\begin{equation}
\left\vert m(t)\right\rangle =e^{-i\frac{\nu_{0}}{4\omega_{0}}\left[
2\omega_{0}t-\sin\left(  2\omega_{0}t\right)  \right]  }\left[  \cos\left(
\omega_{0}t\right)  \left\vert 0\right\rangle +e^{i\nu_{0}t}\sin\left(
\omega_{0}t\right)  \left\vert 1\right\rangle \right]  \text{.}\label{s5a}%
\end{equation}
Then, setting the matrix representation of the Hamiltonian \textrm{H}%
$(t)=i(\left\vert \dot{m}\right\rangle \left\langle m\right\vert -\left\vert
m\right\rangle \left\langle \dot{m}\right\vert )$ in the orthogonal basis
$\left\{  \left\vert m\right\rangle \text{, }\left\vert \dot{m}\right\rangle
\right\}  $ equal to $\mathrm{H}(t)=h_{0}\left(  t\right)  \mathrm{I}%
+\mathbf{h}\left(  t\right)  \cdot\mathbf{\boldsymbol{\sigma}}$, with
$\left\vert m\right\rangle $ in Eq. (\ref{s5a}), and, in addition, writing
$\rho\left(  t\right)  =\left\vert m(t)\right\rangle \left\langle
m(t)\right\vert =(1/2)\left[  \mathrm{I}+\mathbf{a}(t)\cdot
\mathbf{\boldsymbol{\sigma}}\right]  $, we obtain that the Bloch vector
$\mathbf{a}(t)$ and the magnetic field vector $\mathbf{h}\left(  t\right)  $
are given by%
\begin{equation}
\mathbf{a}(t)\overset{\text{def}}{=}\left(
\begin{array}
[c]{c}%
\sin(\theta)\cos\left(  \varphi\right)  \\
\sin(\theta)\sin\left(  \varphi\right)  \\
\cos(\theta)
\end{array}
\right)  =\left(
\begin{array}
[c]{c}%
\sin\left(  2\omega_{0}t\right)  \cos\left(  \nu_{0}t\right)  \\
\sin\left(  \nu_{0}t\right)  \sin\left(  2\omega_{0}t\right)  \\
\cos\left(  2\omega_{0}t\right)
\end{array}
\right)  \text{,}\label{BV}%
\end{equation}
and\textbf{,}%
\begin{equation}
\mathbf{h}\left(  t\right)  =\left(
\begin{array}
[c]{c}%
-\frac{\dot{\varphi}}{2}\cos(\theta)\sin(\theta)\cos\left(  \varphi\right)
-\frac{\dot{\theta}}{2}\sin(\varphi)\\
-\frac{\dot{\varphi}}{2}\cos(\theta)\sin(\theta)\sin\left(  \varphi\right)
+\frac{\dot{\theta}}{2}\cos(\varphi)\\
\frac{\dot{\varphi}}{2}\sin^{2}(\theta)
\end{array}
\right)  =\left(
\begin{array}
[c]{c}%
-\frac{\nu_{0}}{2}\cos\left(  2\omega_{0}t\right)  \sin(2\omega_{0}%
t)\cos\left(  \nu_{0}t\right)  -\omega_{0}\sin\left(  \nu_{0}t\right)  \\
-\frac{\nu_{0}}{2}\cos\left(  2\omega_{0}t\right)  \sin(2\omega_{0}%
t)\sin\left(  \nu_{0}t\right)  +\omega_{0}\cos\left(  \nu_{0}t\right)  \\
\frac{\nu_{0}}{2}\sin^{2}\left(  2\omega_{0}t\right)
\end{array}
\right)  \text{,}\label{m2}%
\end{equation}
respectively, with $h_{0}\left(  t\right)  =0$. The temporal behavior of the
Cartesian components of the magnetic field vector $\mathbf{h=h}_{\perp
}+\mathbf{h}_{\parallel}=\left[  h_{x}\hat{x}+h_{y}\hat{y}\right]  +h_{z}%
\hat{z}$ is displayed in Fig. $1$. As a consistency check, we verified that
the two real three-dimensional vectors $\mathbf{a}(t)$ and $\mathbf{h}\left(
t\right)  $ in Eqs. (\ref{BV}) and (\ref{m2}), respectively, verify the
relation $\mathbf{\dot{a}}(t)=2\mathbf{h}\left(  t\right)  \times
\mathbf{a}(t)$ as correctly expected (for details, see Appendix A). In Fig.
$2$, we illustrate the (nongeodesic) evolution path on the Bloch sphere that
correspond to the Bloch vector $\mathbf{a}(t)$ in Eq. (\ref{BV}) and generated
by the nonstationary Hamiltonian $\mathrm{H}(t)=\mathbf{h}\left(  t\right)
\cdot\mathbf{\boldsymbol{\sigma}}$ with $\mathbf{h}\left(  t\right)  $ in Eq.
(\ref{m2}).

We are now ready to calculate the curvature coefficient $\kappa_{\mathrm{AC}%
}^{2}(t)$ \ of the quantum evolution which drives $\left\vert
m(t)\right\rangle $ in Eq. (\ref{s5a}) under the nonstationary Hamiltonian
$\mathrm{H}(t)=\mathbf{h}\left(  t\right)  \cdot\mathbf{\boldsymbol{\sigma}}$,
with $\mathbf{h}\left(  t\right)  $ in Eq. (\ref{m2}) and with no waste of
energy since $\eta_{\mathrm{SE}}\left(  t\right)  =\sqrt{\left\langle \dot
{m}\left(  t\right)  \left\vert \dot{m}\left(  t\right)  \right.
\right\rangle }/\sqrt{\left\langle \dot{m}\left(  t\right)  \left\vert \dot
{m}\left(  t\right)  \right.  \right\rangle }$ is constantly equal to
one.\begin{figure}[t]
\centering
\includegraphics[width=0.35\textwidth] {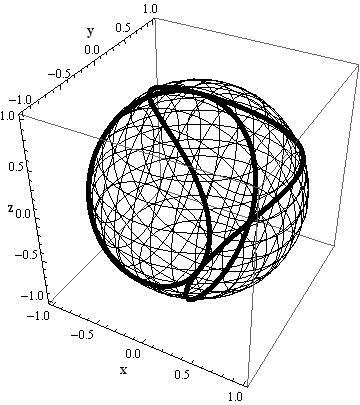}\caption{Illustrative depiction
of nongeodesic evolution paths (solid thick) on the Bloch sphere generated by
the nonstationary Hamiltonian considered in the paper. For simplicity, we set
here $\nu_{0}=\omega_{0}=1$ and $0\leq t\leq20$. Physical units are chosen
using $\hslash=1$.}%
\end{figure}

\section{Quantum evolution}

Before calculating the curvature $\kappa_{\mathrm{AC}}^{2}(t)$, we wish to
study additional aspects of the quantum evolution so that we can gain a
clearer physical interpretation of its curvature.

\emph{Geodesic efficiency}. We begin by observing that the quantum evolution
is nongeodesic. Indeed, consider the evolution between two orthogonal quantum
states $\left\vert m(t_{A})\right\rangle $ and $\left\vert m(t_{B}%
)\right\rangle $ with $t_{A}=0$ and $t_{B}=\pi/(2\omega_{0})$. The energy
dispersion $\Delta E\left(  t\right)  \overset{\text{def}}{=}\sqrt
{\mathrm{tr}\left(  \rho\mathrm{H}^{2}\right)  -\left[  \mathrm{tr}\left(
\rho\mathrm{H}\right)  \right]  ^{2}}=\sqrt{\left\langle \dot{m}(t)\left\vert
\dot{m}(t)\right.  \right\rangle }$ is given by,%
\begin{equation}
\Delta E\left(  t\text{; }\omega_{0}\text{, }\nu_{0}\right)  =\omega_{0}%
\sqrt{1+\frac{1}{4}\left(  \frac{\nu_{0}}{\omega_{0}}\right)  ^{2}\sin
^{2}\left(  2\omega_{0}t\right)  }\text{.}%
\end{equation}
Then, the Anandan-Aharonov geodesic efficiency $\eta_{\mathrm{GE}}$ defined as
\cite{anandan90,cafaro20}%
\begin{equation}
\eta_{\mathrm{GE}}\overset{\text{def}}{=}\frac{2\arccos\left[  \left\vert
\left\langle m(t_{A})\left\vert m(t_{B})\right.  \right\rangle \right\vert
\right]  }{2\int_{t_{A}}^{t_{B}}\frac{\Delta E\left(  t^{\prime}\right)
}{\hslash}dt^{\prime}}\text{,}\label{eta1}%
\end{equation}
becomes%
\begin{equation}
\eta_{\mathrm{GE}}\left(  \omega_{0}\text{, }\nu_{0}\right)  =\frac{\pi}%
{2}\frac{1}{\mathrm{E}\left(  -\frac{1}{4}\left(  \frac{\nu_{0}}{\omega_{0}%
}\right)  ^{2}\right)  }<1\text{,}\label{geoeff}%
\end{equation}
where \textrm{E}$\left(  m\right)  $ in Eq. (\ref{geoeff}) is the complete
elliptic integral of the second kind with parameter $m$ \cite{grad00}. For
instance, setting $\nu_{0}=\omega_{0}=1$, $\mathrm{E}\left(  -1/4\right)
\simeq1.66>\pi/2$. Since $\eta_{\mathrm{GE}}<1$ from Eq. (\ref{geoeff}), we
expect that $\kappa_{\mathrm{AC}}^{2}(t)\neq0$. Interestingly, for $\nu_{0}%
=0$, $\mathrm{E}\left(  0\right)  =\pi/2$ and $\eta_{\mathrm{GE}}$ in Eq.
(\ref{geoeff}) equals one. For this choice of parameters, the state
$\left\vert m(t)\right\rangle =\cos(\omega_{0}t)\left\vert 0\right\rangle
+\sin(\omega_{0}t)\left\vert 1\right\rangle $ describes a great circle on the
Bloch sphere. Thus, the quantum evolution becomes geodesic.

\emph{Speed}. The speed of quantum evolution in projective Hilbert space is
given by $v_{\mathrm{H}}\left(  t\right)  $ with
\begin{equation}
v_{\mathrm{H}}^{2}\left(  t\right)  \overset{\text{def}}{=}\left[  \Delta
E\left(  t\right)  \right]  ^{2}=\left\langle \dot{m}(t)\left\vert \dot
{m}(t)\right.  \right\rangle =\omega_{0}^{2}+\frac{\nu_{0}^{2}}{4}\sin
^{2}(2\omega_{0}t)\text{.}%
\end{equation}
Therefore, $v_{\mathrm{H}}\left(  t\right)  $ equals%
\begin{equation}
v_{\mathrm{H}}\left(  t\text{; }\omega_{0}\text{, }\nu_{0}\right)  =\omega
_{0}\sqrt{1+\frac{1}{4}\left(  \frac{\nu_{0}}{\omega_{0}}\right)  ^{2}\sin
^{2}(2\omega_{0}t)}\text{.} \label{s1}%
\end{equation}
From Eq. (\ref{s1}), we observe that $\bar{v}_{\mathrm{H}}\overset{\text{def}%
}{=}\max_{t}\left[  v_{\mathrm{H}}\left(  t\right)  \right]  =\omega_{0}%
\sqrt{1+(1/4)\left(  \nu_{0}/\omega_{0}\right)  ^{2}}$ is reached at $t_{\max
}^{v}\overset{\text{def}}{=}\pi/(4\omega_{0})+\left[  \pi/(2\omega
_{0})\right]  n$, with $n\in%
\mathbb{Z}
$. In particular, the instances $t_{\max}^{v}$ are obtained by noting that
$\sin^{2}(2\omega_{0}t)=\left[  1-\cos(4\omega_{0}t)\right]  /2=1$ iff
$4\omega_{0}t_{\max}^{v}=\pi+2\pi n$. Similarly, $\underline{v}_{\mathrm{H}%
}\overset{\text{def}}{=}\min_{t}\left[  v_{\mathrm{H}}\left(  t\right)
\right]  =\omega_{0}$ is obtained when $4\omega_{0}t_{\min}^{v}=2\pi n$, that
is for $t_{\min}^{v}\overset{\text{def}}{=}\left[  \pi/(2\omega_{0})\right]
n$, with $n\in%
\mathbb{Z}
$. Finally, observe that $v_{\mathrm{H}}\left(  t\text{; }\omega_{0}\text{,
}\nu_{0}\right)  $ in Eq. (\ref{s1}) is periodic of $T\overset{\text{def}}%
{=}\pi/(2\omega_{0})$.\begin{figure}[t]
\centering
\includegraphics[width=0.85\textwidth] {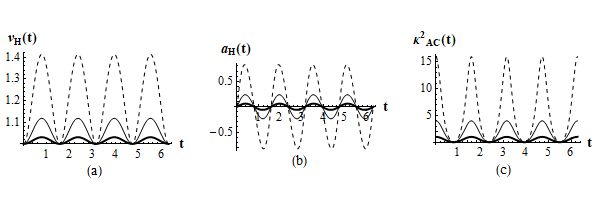}\caption{Graphical visualization
of the quantum evolution speed $v_{\mathrm{H}}(t)$ (a), the acceleration of
the quantum evolution $a_{\mathrm{H}}(t)$ (b) and, finally, the curvature
coefficient $\kappa_{\mathrm{AC}}^{2}(t)$ (c) versus time $t$ with $0\leq
t\leq2\pi$. Dashed, solid thin, and solid thick lines are specified by the
conditions $\nu_{0}=2\omega_{0}$, $\nu_{0}=\omega_{0}$, and $\nu_{0}%
=\omega_{0}/2$, respectively. In all cases, we set $\omega_{0}=1$. Physical
units are chosen so that $\hslash=1$.}%
\end{figure}

\emph{Acceleration}. The acceleration $a_{\mathrm{H}}\left(  t\right)  $ of
the quantum evolution is the rate of change in time of the magnitude of the
speed of quantum evolution $v_{\mathrm{H}}\left(  t\right)  $. Therefore, it
is given by $a_{\mathrm{H}}\left(  t\right)  \overset{\text{def}}%
{=}dv_{\mathrm{H}}\left(  t\right)  /dt$ and, in our case, becomes%
\begin{equation}
a_{\mathrm{H}}\left(  t\text{; }\omega_{0}\text{, }\nu_{0}\right)  =\frac
{1}{4}\frac{\nu_{0}^{2}\sin(4\omega_{0}t)}{\sqrt{1+\frac{1}{4}\left(
\frac{\nu_{0}}{\omega_{0}}\right)  ^{2}\sin^{2}\left(  2\omega_{0}t\right)  }%
}\text{.}\label{a1}%
\end{equation}
Following the reasoning provided for the speed of quantum evolution, we note
that $\bar{a}_{\mathrm{H}}\overset{\text{def}}{=}\max_{t}\left[
a_{\mathrm{H}}\left(  t\right)  \right]  =\left(  \nu_{0}^{2}/4\right)
\left[  1+\frac{1}{8}\left(  \frac{\nu_{0}}{\omega_{0}}\right)  ^{2}\right]
^{-1/2}$ at $t_{\max}^{a}=\pi/(8\omega_{0})+\left[  \pi/(2\omega_{0})\right]
n$, with $n\in%
\mathbb{Z}
$. Furthermore, $\underline{a}_{\mathrm{H}}\overset{\text{def}}{=}\min
_{t}\left[  a_{\mathrm{H}}\left(  t\right)  \right]  =-\bar{a}_{\mathrm{H}}$
at $t_{\min}^{a}=3\pi/(8\omega_{0})+\left[  \pi/(2\omega_{0})\right]  n$, with
$n\in%
\mathbb{Z}
$. Finally, notice that $a_{\mathrm{H}}\left(  t\text{; }\omega_{0}\text{,
}\nu_{0}\right)  $ in Eq. (\ref{a1}) is periodic of $T\overset{\text{def}}%
{=}\pi/(2\omega_{0})$.

\emph{Magnetic field components}. To illustrate how the parameters $\omega
_{0}$ and $\nu_{0}$ are related to the non-static configuration of the
magnetic field, we consider the ratio between the intensity of the magnetic
field component along the $z$-axis (i.e., parallel to the quantization axis)
and the intensity of the magnetic field component that lays in the $xy$-plane
(i.e., transverse to the quantization axis). This consideration is reasonable.
Indeed, for example, in typical experimental conditions for quantum-mechanical
Rabi oscillations, the ratio between these two intensities is much greater
than one \cite{sakurai,rabi54,carlo19,steven20}. Specifically, $\mathbf{h}%
\left(  t\right)  $ in Eq. (\ref{m2}) can be decomposed in a parallel
component $\mathbf{h}_{\parallel}$ along the $z$-axis and a transverse
component $\mathbf{h}_{\perp}$ that is perpendicular to the $z$-axis (i.e., a
component in the $xy$-plane). Specifically, we can write $\mathbf{h}%
\overset{\text{def}}{\mathbf{=}}\mathbf{h}_{\perp}+\mathbf{h}_{\parallel}$,
with $\mathbf{h}_{\perp}\overset{\text{def}}{\mathbf{=}}\left(  \mathbf{h\cdot
}\hat{x}\right)  \hat{x}+\left(  \mathbf{h\cdot}\hat{y}\right)  \hat{y}$ and
$\mathbf{h}_{\parallel}\overset{\text{def}}{\mathbf{=}}\left(  \mathbf{h\cdot
}\hat{z}\right)  \hat{z}$. In our case, we have $\mathbf{h}_{\perp}%
^{2}=\left(  \nu_{0}^{2}/16\right)  \sin^{2}(4\omega_{0}t)+\omega_{0}^{2}$ and
$\mathbf{h}_{\parallel}^{2}=\left(  \nu_{0}^{2}/4\right)  \sin^{4}(2\omega
_{0}t)$. Therefore, the ratio $\left(  \mathbf{h}_{\parallel}^{2}%
/\mathbf{h}_{\perp}^{2}\right)  (t$; $\omega_{0}$, $\nu_{0})$ reduces to%
\begin{equation}
\left(  \frac{\mathbf{h}_{\parallel}^{2}}{\mathbf{h}_{\perp}^{2}}\right)
(t\text{; }\omega_{0}\text{, }\nu_{0})=4\frac{\sin^{4}(2\omega_{0}t)}{\sin
^{2}(4\omega_{0}t)+16\left(  \frac{\omega_{0}}{\nu_{0}}\right)  ^{2}}\text{.}
\label{ratio}%
\end{equation}
From Eq. (\ref{ratio}), we observe that the local maxima of the ratio
$\mathbf{h}_{\parallel}^{2}/\mathbf{h}_{\perp}^{2}$ are given by
$(\mathbf{h}_{\parallel}^{2}/\mathbf{h}_{\perp}^{2})_{\max}=(1/4)(\nu
_{0}/\omega_{0})^{2}$ and are reached at $\overline{t}_{\ast}=\pi/(4\omega
_{0})+\left[  \pi/(2\omega_{0})\right]  n$, with $n\in%
\mathbb{Z}
$. Moreover, the local minima of the ratio $\mathbf{h}_{\parallel}%
^{2}/\mathbf{h}_{\perp}^{2}$ are given by $(\mathbf{h}_{\parallel}%
^{2}/\mathbf{h}_{\perp}^{2})_{\min}=0$ and are reached at $\underline{t}%
_{\ast}=\left[  \pi/(2\omega_{0})\right]  n$ with $n\in%
\mathbb{Z}
$. Finally, observe that $\mathbf{h}_{\parallel}^{2}/\mathbf{h}_{\perp}^{2}$
in Eq. (\ref{ratio}) is periodic of $T\overset{\text{def}}{=}\pi/(2\omega
_{0})$.

\emph{Curvature}. We are now ready to calculate the curvature coefficient
$\kappa_{\mathrm{AC}}^{2}(t)$ \ of the quantum evolution which drives
$\left\vert m(t)\right\rangle $ in Eq. (\ref{s5a}) under the nonstationary
Hamiltonian $\mathrm{H}(t)=\mathbf{h}\left(  t\right)  \cdot
\mathbf{\boldsymbol{\sigma}}$, with $\mathbf{h}\left(  t\right)  $ in Eq.
(\ref{m2}). Then, given $\mathbf{a}\left(  t\right)  $ in Eq. (\ref{BV}) and
$\mathbf{h}(t)$ in Eq. (\ref{m2}), we note that $\mathbf{a}\cdot\mathbf{h}=0$
and $\mathbf{a\cdot\dot{h}}=0$. For completeness, we remark that no tedious
calculation is necessary to explain that $\mathbf{a\cdot\dot{h}}=0$ once one
verifies that $\mathbf{a}\cdot\mathbf{h}=0$. Indeed, exploiting the constraint
equation $\mathbf{\dot{a}=2h\times a}$ (for details, see Appendix A),
$\mathbf{a\cdot\dot{h}}=0$ follows from elementary algebraic steps and simple
geometric reasoning. Therefore, the curvature coefficient $\kappa
_{\mathrm{AC}}^{2}(\mathbf{a}$, $\mathbf{h)}$ in Eq. (\ref{XXX}) reduces to%
\begin{equation}
\kappa_{\mathrm{AC}}^{2}(\mathbf{a}\text{, }\mathbf{h)}=\frac{\mathbf{h}%
^{2}\mathbf{\dot{h}}^{2}-\left(  \mathbf{h\cdot\dot{h}}\right)  ^{2}%
}{\mathbf{h}^{6}}\text{,}\label{u21}%
\end{equation}
that is, after some algebra, we get that%
\begin{equation}
\kappa_{\mathrm{AC}}^{2}(t\text{; }\omega_{0}\text{, }\nu_{0})=\frac{\sin
^{2}\left(  4\omega_{0}t\right)  +32(\frac{\omega_{0}}{\nu_{0}})^{2}\left[
1+\cos\left(  4\omega_{0}t\right)  \right]  }{\left[  \sin^{2}\left(
2\omega_{0}t\right)  +4(\frac{\omega_{0}}{\nu_{0}})^{2}\right]  ^{2}}%
-4(\frac{\omega_{0}}{\nu_{0}})^{2}\frac{\sin^{2}\left(  4\omega_{0}t\right)
}{\left[  \sin^{2}\left(  2\omega_{0}t\right)  +4(\frac{\omega_{0}}{\nu_{0}%
})^{2}\right]  ^{3}}\text{.}\label{curvature}%
\end{equation}
It is worthwhile pointing out that, unlike what happens in quantum evolutions
governed by stationary Hamiltonians \cite{alsing24A}, the condition
$\mathbf{a}\cdot\mathbf{h}=0$ is not sufficient for having a vanishing
curvature coefficient. As a side remark, we stress that this condition becomes
sufficient only if we add the additional assumption that the magnetic field
vector does not change direction in time (i.e., if $\mathbf{h}$ and
$\mathbf{\dot{h}}$ are collinear). Unfortunately, as evident from Eq.
(\ref{m2}), this is not the case in the problem being analyzed here. We
emphasize that $\left(  \mathrm{\kappa}_{\mathrm{AC}}^{2}\right)  _{\max
}\overset{\text{def}}{=}\max_{t}\left[  \kappa_{\mathrm{AC}}^{2}\left(
t\right)  \right]  =4(\nu_{0}/\omega_{0})^{2}$ at $t_{\max}^{\mathrm{\kappa
}^{2}}=\left[  \pi/(2\omega_{0})\right]  n$, with $n\in%
\mathbb{Z}
$. Furthermore, $\left(  \mathrm{\kappa}_{\mathrm{AC}}^{2}\right)  _{\min
}\overset{\text{def}}{=}\min_{t}\left[  \kappa_{\mathrm{AC}}^{2}\left(
t\right)  \right]  =0$ at $t_{\min}^{\mathrm{\kappa}^{2}}=\pi/(4\omega
_{0})+\left[  \pi/(2\omega_{0})\right]  n$, with $n\in%
\mathbb{Z}
$. Interestingly, as $\nu_{0}$ approaches zero, we correctly recover a
geodesic quantum evolution with a vanishing curvature coefficient
$\kappa_{\mathrm{AC}}^{2}$ in Eq. (\ref{curvature}). Finally, notice that
$\kappa_{\mathrm{AC}}^{2}(t$; $\omega_{0}$, $\nu_{0})$ in Eq. (\ref{curvature}%
) is periodic of $T\overset{\text{def}}{=}\pi/(2\omega_{0})$. To the best of
our knowledge, Eq. (\ref{curvature}) is the first example of a closed form
expression for the curvature coefficient of a quantum evolution specified by a
two-parameter nonstationary (traceless) Hamiltonian for two-level quantum systems.

In what follows, we summarize a number of considerations on the temporal
behaviors of $v_{\mathrm{H}}(t)$ in Eq. (\ref{s1}), $a_{\mathrm{H}}(t)$ in Eq.
(\ref{a1}), $(\mathbf{h}_{\parallel}^{2}/\mathbf{h}_{\perp}^{2})\left(
t\right)  $ in Eq. (\ref{ratio}), and $\kappa_{\mathrm{AC}}^{2}\left(
t\right)  $ in Eq. (\ref{curvature}). First, each and everyone of these
quantities displays a periodic oscillatory behavior characterized by a period
$T\overset{\text{def}}{=}\pi/(2\omega_{0})$. Second, the relation between the
behaviors of $v_{\mathrm{H}}(t)$ and $a_{\mathrm{H}}(t)\overset{\text{def}}%
{=}dv_{\mathrm{H}}(t)/dt$ is clear, since the acceleration $a_{\mathrm{H}}(t)$
is the temporal rate of change of the speed $v_{\mathrm{H}}(t)$ of the quantum
evolution in projective Hilbert space. Third, whenever $v_{\mathrm{H}}(t)$
decreases, $\kappa_{\mathrm{AC}}^{2}\left(  t\right)  $ increases (and vice
versa). In particular, whenever $v_{\mathrm{H}}(t)$ reaches its maximum value
(minimum value, respectively), the curvature coefficient $\kappa_{\mathrm{AC}%
}^{2}\left(  t\right)  $ assumes its minimum value (maximum value,
respectively). Fourth, the acceleration $a_{\mathrm{H}}(t)$ vanishes whenever
$\kappa_{\mathrm{AC}}^{2}\left(  t\right)  $ reaches its local minima and
maxima. Fifth, $v_{\mathrm{H}}(t)$ and $(\mathbf{h}_{\parallel}^{2}%
/\mathbf{h}_{\perp}^{2})\left(  t\right)  $ exhibit the same monotonic
behavior. In particular, whenever $v_{\mathrm{H}}(t)$ assumes it maximum value
(minimum value, respectively), $(\mathbf{h}_{\parallel}^{2}/\mathbf{h}_{\perp
}^{2})\left(  t\right)  $ reaches its maximum value (minimum value,
respectively). Finally, given the links between the pairs $\left(
v_{\mathrm{H}}(t)\text{, }\kappa_{\mathrm{AC}}^{2}\left(  t\right)  \right)  $
and $\left(  v_{\mathrm{H}}(t)\text{, }(\mathbf{h}_{\parallel}^{2}%
/\mathbf{h}_{\perp}^{2})\left(  t\right)  \right)  $, the comparative analysis
between the temporal behaviors of $\kappa_{\mathrm{AC}}^{2}\left(  t\right)  $
and $(\mathbf{h}_{\parallel}^{2}/\mathbf{h}_{\perp}^{2})\left(  t\right)  $
becomes self-evident. Finally, while the plots of $v_{\mathrm{H}}(t)$,
$a_{\mathrm{H}}(t)$, $\kappa_{\mathrm{AC}}^{2}\left(  t\right)  $ as functions
of time are illustrated in Fig. $3$ for different choices of the two
parameters $\omega_{0}$ and $\nu_{0}$, the temporal behaviors of
$(\mathbf{h}_{\parallel}^{2}/\mathbf{h}_{\perp}^{2})\left(  t\right)  $ and
$\kappa_{\mathrm{AC}}^{2}\left(  t\right)  $ are displayed in Fig. $4$.

We are now ready for our summary of results and final
comments.\begin{figure}[t]
\centering
\includegraphics[width=1\textwidth] {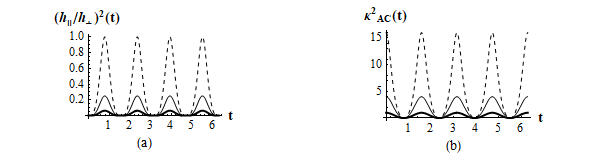}\caption{Plots of the ratio squared
between the parallel and the transverse magnetic field vector intensities
$(h_{\parallel}/h_{\perp})^{2}(t)$ (a) and the curvature coefficient
$\kappa_{\mathrm{AC}}^{2}(t)$ (b) versus time $t$ with $0\leq t\leq2\pi$.
Dashed, solid thin, and solid thick lines are defined by the relations
$\nu_{0}=2\omega_{0}$, $\nu_{0}=\omega_{0}$, and $\nu_{0}=\omega_{0}/2$,
respectively. In all cases, we put $\omega_{0}=1$. Physical units are selected
so that $\hslash=1$.}%
\end{figure}

\section{Concluding remarks}

In this paper, we discussed a first important example of an exact analytical
expression of the curvature coefficient $\kappa_{\mathrm{AC}}^{2}(t$;
$\omega_{0}$, $\nu_{0})$ (Eq. (\ref{curvature}) and Fig. $1$) of a quantum
evolution for a two-level quantum system in a time-dependent magnetic field
$\mathbf{h}\left(  t\right)  $ (Eq. (\ref{m2})). Specifically, we investigated
the dynamics generated by a two-parameter (i.e., $\omega_{0}$ and $\nu_{0}$ in
Eq. (\ref{s5a})) nonstationary Hermitian Hamiltonian $\mathrm{H}%
(t)=\mathbf{h}\left(  t\right)  \cdot\mathbf{\boldsymbol{\sigma}}$ with unit
speed efficiency $\eta_{\mathrm{SE}}$ (Eq. (\ref{se2})). The two parameters
$\omega_{0}$ and $\nu_{0}$ specify the constant temporal rates of change of
the polar and azimuthal angles (i.e., $\omega_{0}=\dot{\theta}/2$ and $\nu
_{0}=\dot{\varphi}$) employed in the Bloch sphere representation of the
evolving pure state (with Bloch vector $\mathbf{a}\left(  t\right)  $ in Eq.
(\ref{BV}) and in Fig. $2$). To better grasp the physical significance of the
curvature coefficient $\kappa_{\mathrm{AC}}^{2}(t$; $\omega_{0}$, $\nu_{0})$,
showing that the quantum evolution is nongeodesic since the geodesic
efficiency $\eta_{\mathrm{GE}}$ of the quantum evolution is strictly less than
one (Eq. (\ref{geoeff})) and tuning the two Hamiltonian parameters, we
compared in Fig. $3$ the temporal behavior of the curvature coefficient
$\kappa_{\mathrm{AC}}^{2}(t$; $\omega_{0}$, $\nu_{0})$ with that of the speed
$v_{\mathrm{H}}(t)$ (\ref{s1}) and the acceleration $a_{\mathrm{H}}(t)$
(\ref{a1}) of the evolution in projective Hilbert space. Furthermore, we
compared in Fig. $4$ the temporal profile of the curvature coefficient
$\kappa_{\mathrm{AC}}^{2}(t$; $\omega_{0}$, $\nu_{0})$ with that of the square
of the ratio $(\mathbf{h}_{\parallel}^{2}/\mathbf{h}_{\perp}^{2})(t)$ between
the parallel and transverse magnetic field intensities (Eq. (\ref{ratio})).

\medskip

The primary conclusions to be drawn can be outlined as follows. First, unlike
what happens in a stationary setting, the condition $\mathbf{a}\cdot
\mathbf{h=}0$ is neither necessary nor sufficient to get a quantum evolution
with a vanishing curvature in the time-varying framework. Instead, when
$\mathbf{h}$ does not change in direction (i.e., if $\mathbf{h}$ and
$\mathbf{\dot{h}}$ are collinear) and $\mathbf{a}\cdot\mathbf{h=}0$, the
curvature vanishes. Second, the bending of quantum paths on the Bloch sphere
does not need to happen with waste of energy resources (i.e., $\kappa
_{\mathrm{AC}}^{2}\neq0$ does not imply $\eta_{\mathrm{SE}}<1$). Third, no
bending happens only with unit geodesic efficiency (i.e., $\kappa
_{\mathrm{AC}}^{2}=0$ only when $\eta_{\mathrm{GE}}=1$). More specifically,
our notion of bending is defined in terms of our proposed curvature
coefficient $\kappa_{\mathrm{AC}}^{2}$. For a geodesic motion on the Bloch
sphere equipped with the Fubini-Study metric, $\kappa_{\mathrm{AC}}^{2}$
vanishes and the geodesic efficiency\textbf{ }$\eta_{\mathrm{GE}}$ equals one.
Instead, the bending quantified by the Frenet-Serret curvature coefficient
would be always present when lines are not straight (for example, consider a
circle in the equatorial plane of a sphere in $%
\mathbb{R}
^{3}$).\textbf{ }Moreover, despite the fact that the bending phenomenon can
emerge, in principle, from the (deterministic) evolution of an arbitrary
quantum system in a pure state that evolves unitarily according to the
Schr\"{o}dinger equation, its quantification is intrinsically linked to the
probabilistic nature of quantum theory. Indeed, for a discussion on how to
express our curvature and the torsion coefficients of a quantum evolution in
terms of statistical concepts, including kurtosis and skewness, we suggest
Refs. \cite{alsing24A,alsing24B}. Fourth,\textbf{ }regions of high (low)
curvature coefficients $\kappa_{\mathrm{AC}}^{2}$ are characterized by low
(high) speed $v_{\mathrm{H}}$ of quantum evolutions. In particular, the
acceleration $a_{\mathrm{H}}$ of quantum evolutions vanishes when the
curvature coefficient $\kappa_{\mathrm{AC}}^{2}$ achieves its local extrema
(i.e., local minima and local maxima). Fifth, the curvature coefficient
$\kappa_{\mathrm{AC}}^{2}$ and the ratio squared between the parallel and
transverse magnetic field intensities exhibit the same oscillatory behavior
for the scenario being\textbf{ }investigated here. In particular, high (low)
curvature regions correspond to high (low) $\mathbf{h}_{\parallel}%
^{2}/\mathbf{h}_{\perp}^{2}$ ratios.

\medskip

As pointed out in Ref. \cite{laba17}, the geometric concepts of curvature and
torsion of quantum evolutions are interesting on their own rights.
Unfortunately, these notions were only defined for stationary Hamiltonians in
Ref. \cite{laba17}. In recent years, being their applicability limited to
time-independent Hamiltonian settings, these concepts have been successfully
used to characterize geometric aspects of graph states in Ref.
\cite{gnatenko22} and geometric features of unitary dynamical evolutions
employed to solve combinatorial optimization problems with continuous-time
quantum algorithms in Ref. \cite{banks24}. Our geometric analysis, instead,
paves the way to fully time-dependent curvature investigations of quantum
evolutions that go beyond the very-short-time limits of the expectation values
of suitable observable of the quantum systems being analyzed \cite{banks24}.

\medskip

As mentioned in the Introduction, the notions of curvature and torsion of
quantum evolutions have been remarkably reconstructed in an alternative
fashion (with respect to Ref. \cite{laba17}) in a time-independent setting in
Ref. \cite{alsing24A} and, most importantly, extended to a fully
time-dependent setting in Ref. \cite{alsing24B}. However, the illustrative
single-qubit example discussing the concept of curvature in Ref.
\cite{alsing24B} was carried out in a numerical fashion. Furthermore, no
discussion on the link between curvature and other quantum evolution
quantifiers such as the speed efficiency, the geodesic efficiency, and the
magnetic field parallel and transverse configurations was presented in Ref.
\cite{alsing24B}. In view of these considerations, despite the fact that our
analytical investigation is limited to a two-level quantum system (and, in
addition, to a particular time-varying magnetic field configuration) and its
extension to higher-dimensional systems is expected to be nontrivial
\cite{alsing24B}, we believe that our analysis constitutes a nontrivial piece
of work that paves the way to geometric characterizations of more realistic
quantum evolutions.

\medskip

We remark that, modulo computational tediousness, the main difficulties that
emerge when applying our approach to arbitrary nonstationary magnetic field
configurations for two-level quantum systems arise from the complications in
obtaining exact analytical solutions to the time-dependent Schr\"{o}dinger
equation
\cite{landau32,zener32,rabi37,barnes12,barnes13,messina14,grimaudo18,elena20,grimaudo23}%
. Moreover, while the formula for $\kappa_{\mathrm{AC}}^{2}$ is theoretically
valid for any $d$-level quantum system that evolves under a nonstationary
Hamiltonian, our study was confined to systems with just two levels. In this
more straightforward situation, the understanding and visual representation of
Bloch vectors and Bloch spheres are straightforward, unlike in more complex,
higher-dimensional situations. As we move from systems with just two levels to
those with more dimensions, the clarity of these visualizations decreases. In
fact, quantum theory exhibits unique characteristics in more complex systems,
including the simplest yet non-trivial example with three levels, which are
known as qutrits \cite{kurzy11}. These peculiar quantum behaviors make it
challenging to understand the geometric structure of quantum systems in higher
dimensions \cite{xie20,siewert21}. We recommend Ref. \cite{gamel16} for a
comprehensive overview of how Bloch vectors are used to represent
single-qubit, single-qutrit, and two-qubit systems, using matrices such as
Pauli, Gell-Mann, and Dirac matrices, respectively\textbf{.}

\textbf{\medskip}

Realistic evolutions are expected to occur in larger Hilbert spaces where the
emergence of quantum entanglement and, in addition, the nature of the possibly
time-dependent interactions among the various elements of the composite system
can lead to a rather rich dynamical setting where geometric concepts like
curvature and torsion can be a helpful investigative tool
\cite{gnatenko22,banks24}. Moreover, given the intriguing link between the
concepts of complexity and efficiency of quantum evolutions
\cite{cafaroPRE20,cafaroPRE22,cafaroPRD22} together with the connection
between curvature and efficiencies mentioned here, we expect the concept of
curvature to play a determinant role in determining the complexity of
quantum-mechanical evolutions
\cite{leo17,chapman18,vijay20,ali20,iaconis21,vijay22,caputa22} as well.

\medskip

To sum up, even though it has its present constraints, we truly believe that
our research will encourage other researchers and set the stage for more
in-depth studies on how geometry and quantum mechanics interact.

\begin{acknowledgments}
C.C. thanks Christian Corda for helpful remarks. Furthermore, the authors are
grateful to anonymous referees for constructive comments leading to an
improved version of the paper.\textbf{ }Any opinions, findings and conclusions
or recommendations expressed in this material are those of the author(s) and
do not necessarily reflect the views of their home Institutions.
\end{acknowledgments}

\appendix

\section{Derivation of $\mathbf{\dot{a}=}2\mathbf{h\times a}$}

In this Appendix, we present two different derivations of the condition
$\mathbf{\dot{a}=}2\mathbf{h\times a}$ mentioned in Sections III and IV. The
first derivation is based on formal quantum-mechanical rules. The second one,
instead, is a rederivation of the original one proposed by Feynman and
collaborators in Ref. \cite{feynman57}.

\subsection{Abstract derivation}

We begin by deriving the relation $\mathbf{\dot{a}=}2\mathbf{h\times a}$ with
formal quantum-mechanical methods. Consider a nonstationary Hamiltonian
defined as \textrm{H}$\left(  t\right)  \overset{\text{def}}{=}\mathbf{h}%
\left(  t\right)  \cdot\mathbf{\boldsymbol{\sigma}}$, while the pure state is
given by $\rho\left(  t\right)  =\left\vert \psi\left(  t\right)
\right\rangle \left\langle \psi\left(  t\right)  \right\vert =\left[
\mathrm{I}+\mathbf{a}\left(  t\right)  \cdot\mathbf{\boldsymbol{\sigma}%
}\right]  /2$. Recall that the expectation value $\left\langle Q\right\rangle
_{\rho}$ of an arbitrary qubit observable $Q\overset{\text{def}}{=}%
q_{0}\mathrm{I}+\mathbf{q\cdot\boldsymbol{\sigma}}$ with $q_{0}\in%
\mathbb{R}
$ and $\mathbf{q\in%
\mathbb{R}
}^{3}$ is given by $\left\langle Q\right\rangle _{\rho}=q_{0}+\mathbf{a\cdot
q}$. Then, the expectation value of the time derivative of the Hamiltonian
operator reduces to%
\begin{equation}
\left\langle \mathrm{\dot{H}}\right\rangle =\mathrm{tr}\left[  \rho
\mathrm{\dot{H}}\right]  =\mathbf{a\cdot\dot{h}}\text{.}\label{yo1}%
\end{equation}
Furthermore, making use of standard quantum mechanics rules, we get
\begin{equation}
\left\langle \mathrm{\dot{H}}\right\rangle =\partial_{t}\left\langle
\mathrm{H}\right\rangle =\partial_{t}\left(  \mathbf{a\cdot h}\right)
=\mathbf{\dot{a}\cdot h+a\cdot\dot{h}}\text{.}\label{yo2}%
\end{equation}
From Eqs. (\ref{yo1}) and (\ref{yo2}), there appears to be a potential
incompatibility. Luckily, this is a false suspicion given that $\mathbf{\dot
{a}\cdot h}=0$ since one can demonstrate that $\mathbf{\dot{a}=}%
2\mathbf{h\times a}$ and, thus, $\mathbf{\dot{a}}$ is orthogonal to
$\mathbf{h}$ (and to $\mathbf{a}$) so that $\mathbf{\dot{a}\cdot h=}0$.
Therefore, we have to verify that $\mathbf{\dot{a}=}2\mathbf{h\times a}$. We
start by noting that $\left\langle \mathbf{\boldsymbol{\sigma}}\right\rangle $
is equal to,
\begin{equation}
\left\langle \mathbf{\boldsymbol{\sigma}}\right\rangle =\mathrm{tr}\left[
\rho\mathbf{\boldsymbol{\sigma}}\right]  =\frac{1}{2}\mathrm{tr}\left[
\left(  \mathbf{a\cdot\boldsymbol{\sigma}}\right)  \mathbf{\boldsymbol{\sigma
}}\right]  =\mathbf{a}\text{.}\label{sware}%
\end{equation}
Therefore, employing the Schr\"{o}dinger evolution equation $i\hslash
\partial_{t}\left\vert \psi\left(  t\right)  \right\rangle =\mathrm{H}\left(
t\right)  \left\vert \psi\left(  t\right)  \right\rangle $ with $\hslash=1$,
the time derivative of $\left\langle \mathbf{\boldsymbol{\sigma}}\right\rangle
$ in\ Eq. (\ref{sware}) becomes%
\begin{align}
\mathbf{\dot{a}} &  =\partial_{t}\left\langle \mathbf{\boldsymbol{\sigma}%
}\right\rangle \nonumber\\
&  =\partial_{t}\left(  \left\langle \psi\left\vert \mathbf{\boldsymbol{\sigma
}}\right\vert \psi\right\rangle \right)  \nonumber\\
&  =\left\langle \dot{\psi}\left\vert \mathbf{\boldsymbol{\sigma}}\right\vert
\psi\right\rangle +\left\langle \psi\left\vert \mathbf{\boldsymbol{\sigma}%
}\right\vert \dot{\psi}\right\rangle \nonumber\\
&  =i\left\langle \psi\left\vert \mathrm{H}\mathbf{\boldsymbol{\sigma}%
}\right\vert \psi\right\rangle -i\left\langle \psi\left\vert
\mathbf{\boldsymbol{\sigma}}\mathrm{H}\right\vert \dot{\psi}\right\rangle
\nonumber\\
&  =i\left\langle \psi\left\vert \left[  \mathrm{H}\text{, }%
\mathbf{\boldsymbol{\sigma}}\right]  \right\vert \psi\right\rangle \text{,}%
\end{align}
that is, the time derivative $\mathbf{\dot{a}}$ of the nonstationary Bloch
vector $\mathbf{a}$ reduces to%
\begin{equation}
\mathbf{\dot{a}}=i\left\langle \psi\left\vert \left[  \mathrm{H}\text{,
}\mathbf{\boldsymbol{\sigma}}\right]  \right\vert \psi\right\rangle
\text{.}\label{a-dot}%
\end{equation}
To further simplify Eq. (\ref{a-dot}), we note that the commutator $\left[
\mathrm{H}\text{, }\sigma_{j}\right]  $ can be recast as
\begin{align}
\left[  \mathrm{H}\text{, }\sigma_{j}\right]   &  =\left[  \mathbf{h}%
\cdot\mathbf{\boldsymbol{\sigma}}\text{, }\sigma_{j}\right]  \nonumber\\
&  =\left[  h_{i}\sigma_{i}\text{, }\sigma_{j}\right]  \nonumber\\
&  =h_{i}\left[  \sigma_{i}\text{, }\sigma_{j}\right]  \nonumber\\
&  =h_{i}\left(  2i\epsilon_{ijk}\sigma_{k}\right)  \nonumber\\
&  =-2i\epsilon_{ikj}h_{i}\sigma_{k}\text{,}%
\end{align}
that is,
\begin{equation}
\left[  \mathrm{H}\text{, }\sigma_{j}\right]  =-2i\epsilon_{ikj}h_{i}%
\sigma_{k}\text{,}\label{b-dot}%
\end{equation}
for any $1\leq j\leq3$. Therefore, from Eqs. (\ref{a-dot}) and (\ref{b-dot}),
we obtain%
\begin{align}
\left(  \mathbf{\dot{a}}\right)  _{j} &  \mathbf{=}i\left(  -2i\epsilon
_{ikj}h_{i}\left\langle \sigma_{k}\right\rangle \right)  \nonumber\\
&  =2\epsilon_{ikj}h_{i}a_{k}\nonumber\\
&  =2\left(  \mathbf{h\times a}\right)  _{j}\text{,}%
\end{align}
that is, we arrive at the relation $\mathbf{\dot{a}=}2\mathbf{h\times a}$.
This equation expresses the fact that the Bloch vector $\mathbf{a}(t)$ rotates
about the instantaneous \textquotedblleft magnetic field\textquotedblright%
\ $\mathbf{h}(t)$ defined by the Hamiltonian $\mathrm{H}(t)\overset
{\text{def}}{=}\mathbf{h}(t)\cdot\mathbf{\boldsymbol{\sigma}}$ at each instant
of time $t$. Finally, we end by observing that $\partial_{t}\left(
\mathbf{a\cdot h}\right)  =\mathbf{\dot{a}\cdot h+a\cdot\dot{h}=a\cdot\dot{h}%
}$ since $\mathbf{\dot{a}\cdot h=}0$ because $\mathbf{\dot{a}=}%
2\mathbf{h\times a}$ is perpendicular to the vector $\mathbf{h}$.

In what follows, we present an alternative derivation of the condition
$\mathbf{\dot{a}=}2\mathbf{h\times a}$ inspired by the original calculation by
Feynman and collaborators in Ref. \cite{feynman57}.

\subsection{Feynman's derivation}

In Ref. \cite{feynman57}, Feynman and collaborators showed how the dynamics of
a two-level quantum system in a time-dependent background can be geometrically
described in terms of two real three-dimensional vectors, the unit Bloch
vector $\mathbf{r}\left(  t\right)  $ and a vector $\mathbf{\Omega}\left(
t\right)  $ that specifies the perturbation. More specifically, consider a
two-level quantum system whose evolution is governed by the Schr\"{o}dinger
equation given by,%
\begin{equation}
i\hslash\partial_{t}\left\vert \psi\right\rangle =\left(  \mathrm{H}%
_{0}+V\right)  \left\vert \psi\right\rangle \text{,} \label{SE}%
\end{equation}
with $\mathrm{H}_{0}\overset{\text{def}}{=}(\hslash\omega_{0}/2)\sigma_{z}$
and $V=V(t)$ being a time-dependent perturbation. Observe that the full
Hamiltonian \textrm{H}$(t)\overset{\text{def}}{=}\mathrm{H}_{0}+V\left(
t\right)  $ can also be recast as \textrm{H}$(t)\overset{\text{def}}%
{=}(\hslash/2)\mathbf{\Omega}\left(  t\right)  \cdot\mathbf{\boldsymbol{\sigma
}}$. Assume that $\left\vert \psi\right\rangle \overset{\text{def}}{=}a\left(
t\right)  \left\vert \psi_{a}\right\rangle +b(t)\left\vert \psi_{b}%
\right\rangle $ with $\left\{  \left\vert \psi_{a}\right\rangle \text{,
}\left\vert \psi_{b}\right\rangle \right\}  $ being an orthonormal basis for
the Hilbert space of single-qubit quantum states, where $a\left(  t\right)  $
and $b(t)$ are complex probability amplitudes satisfying the normalization
condition $\left\vert a\left(  t\right)  \right\vert ^{2}+\left\vert
b(t)\right\vert ^{2}=1$. In matrix representation, the Schr\"{o}dinger
equation in Eq. (\ref{SE}) becomes,%
\begin{equation}
\left(
\begin{array}
[c]{c}%
i\hslash\frac{da}{dt}\\
i\hslash\frac{db}{dt}%
\end{array}
\right)  =\left(
\begin{array}
[c]{cc}%
\frac{\hslash\omega_{0}}{2}+V_{aa} & V_{ab}\\
V_{ba} & -\frac{\hslash\omega_{0}}{2}+V_{bb}%
\end{array}
\right)  \left(
\begin{array}
[c]{c}%
a\left(  t\right) \\
b\left(  t\right)
\end{array}
\right)  \text{,} \label{uno}%
\end{equation}
where $V_{ij}\left(  t\right)  \overset{\text{def}}{=}\left\langle \psi
_{i}\left\vert V\left(  t\right)  \right\vert \psi_{j}\right\rangle $,
$\left\langle \psi_{i}\left\vert \psi_{j}\right.  \right\rangle =\delta_{ij}$,
and $i$, $j\in\left\{  a\text{, }b\right\}  $. Therefore, the four scalar
equations emerging from\ Eq. (\ref{uno}) are%
\begin{align}
i\hslash\frac{da}{dt}  &  =\left(  \frac{\hslash\omega_{0}}{2}+V_{aa}\right)
a(t)+V_{ab}b(t)\text{, }\nonumber\\
i\hslash\frac{db}{dt}  &  =V_{ba}a\left(  t\right)  +\left(  -\frac
{\hslash\omega_{0}}{2}+V_{bb}\right)  b(t)\text{,}\nonumber\\
-i\hslash\frac{da^{\ast}}{dt}  &  =\left(  \frac{\hslash\omega_{0}}{2}%
+V_{aa}\right)  a^{\ast}\left(  t\right)  +V_{ba}b^{\ast}(t)\text{,
}\nonumber\\
-i\hslash\frac{db^{\ast}}{dt}  &  =V_{ab}a^{\ast}\left(  t\right)  +\left(
-\frac{\hslash\omega_{0}}{2}+V_{bb}\right)  b^{\ast}(t)\text{.} \label{SE2}%
\end{align}
For later use, we also remark that from $(\hslash/2)\mathbf{\Omega}\left(
t\right)  \cdot\mathbf{\boldsymbol{\sigma}}=\mathrm{H}_{0}+V\left(  t\right)
$ and Eq. (\ref{uno}), we have%
\begin{equation}
\mathbf{\Omega}\overset{\text{def}}{=}\left(  \frac{V_{ab}+V_{ba}}{\hslash
}\text{, }i\frac{V_{ab}-V_{ba}}{\hslash}\text{, }\omega_{0}+\frac
{V_{aa}-V_{bb}}{\hslash}\right)  \text{.} \label{omega}%
\end{equation}
In the majority of cases of interest, $V_{aa}=V_{bb}=0$ or, alternatively,
$V_{aa}$ and $V_{bb}$ are negligible compared to $\left(  \hslash/2\right)
\omega_{0}$. Remarkably, putting $\rho\left(  t\right)  =\left\vert
\psi\left(  t\right)  \right\rangle \left\langle \psi\left(  t\right)
\right\vert =\left[  \mathrm{I}+\mathbf{r}\left(  t\right)  \cdot
\mathbf{\boldsymbol{\sigma}}\right]  /2$, Feynman and collaborators were the
first to point out that Eq. (\ref{SE}) can be geometrically recast as%
\begin{equation}
\mathbf{\dot{r}}=\mathbf{\Omega}\times\mathbf{r}\text{,}%
\end{equation}
that is, as%
\begin{equation}
\frac{dx}{dt}=\Omega_{y}z-\Omega_{z}y\text{, }\frac{dy}{dt}=\Omega_{z}%
x-\Omega_{x}z\text{, }\frac{dz}{dt}=\Omega_{x}y-\Omega_{y}x\text{,}
\label{relation}%
\end{equation}
where $\mathbf{r=}\left(  x\text{, }y\text{, }z\right)  $ is the unit Bloch
vector. The real three-dimensional vector $\mathbf{r}$ is defined as%
\begin{equation}
\mathbf{r}\overset{\text{def}}{=}(ab^{\ast}+ba^{\ast}\text{, }i(ab^{\ast
}-ba^{\ast})\text{, }\left\vert a\right\vert ^{2}-\left\vert b\right\vert
^{2})\text{,} \label{r}%
\end{equation}
while $\mathbf{\Omega}$ is given in Eq. (\ref{omega}). From Eq. (\ref{r}), one
can explicitly verify that $\mathbf{r\cdot r=}1$ thanks to the fact that
$\left\vert a\left(  t\right)  \right\vert ^{2}+\left\vert b(t)\right\vert
^{2}=1$. Moreover, given the Hermiticity of the perturbation operator $V(t)$,
one also has that $V_{aa}=V_{aa}^{\ast}$, $V_{ab}=V_{ba}^{\ast}$,
$V_{ba}=V_{ab}^{\ast}$, and $V_{bb}=V_{bb}^{\ast}$. For completeness, we check
here the first relation in Eq. (\ref{relation}). Specifically, we wish to
verify that $dx/dt=\Omega_{y}z-\Omega_{z}y$, that is%
\begin{equation}
\frac{d\left(  ab^{\ast}+ba^{\ast}\right)  }{dt}=i\frac{V_{ab}-V_{ba}}%
{\hslash}\left(  \left\vert a\right\vert ^{2}-\left\vert b\right\vert
^{2}\right)  -i\left(  \omega_{0}+\frac{V_{aa}-V_{bb}}{\hslash}\right)
\left(  ab^{\ast}-ba^{\ast}\right)  \text{.} \label{down}%
\end{equation}
Note that Eq. (\ref{down}) is correct if and only if,%
\begin{align}
-i\hslash\frac{d\left(  ab^{\ast}+ba^{\ast}\right)  }{dt}  &  =-i\hslash
\left(  \frac{da}{dt}b^{\ast}+a\frac{db^{\ast}}{dt}+\frac{db}{dt}a^{\ast
}+b\frac{da^{\ast}}{dt}\right) \nonumber\\
&  =\left(  V_{ab}-V_{ba}\right)  \left(  \left\vert a\right\vert
^{2}-\left\vert b\right\vert ^{2}\right)  -\left(  \hslash\omega_{0}%
+V_{aa}-V_{bb}\right)  \left(  ab^{\ast}-ba^{\ast}\right)  \text{.} \label{up}%
\end{align}
At this point, note that%
\begin{align}
-i\hslash\frac{d\left(  ab^{\ast}+ba^{\ast}\right)  }{dt}  &  =-i\hslash
b^{\ast}\left(  \frac{da}{dt}\right)  -i\hslash a\left(  \frac{db^{\ast}}%
{dt}\right)  -i\hslash a^{\ast}\left(  \frac{db}{dt}\right)  -i\hslash
b\left(  \frac{da^{\ast}}{dt}\right) \nonumber\\
&  =-b^{\ast}\left(  i\hslash\frac{da}{dt}\right)  +a\left(  -i\hslash
\frac{db^{\ast}}{dt}\right)  -a^{\ast}\left(  i\hslash\frac{db}{dt}\right)
+b\left(  -i\hslash\frac{da^{\ast}}{dt}\right) \nonumber\\
&  =-b^{\ast}\left[  \left(  \frac{\hslash\omega_{0}}{2}+V_{aa}\right)
a+V_{ab}b\right]  +a\left[  V_{ab}a^{\ast}+(-\frac{\hslash\omega_{0}}%
{2}+V_{bb})b^{\ast}\right]  +\nonumber\\
&  -a^{\ast}\left[  V_{ba}a+\left(  -\frac{\hslash\omega_{0}}{2}%
+V_{bb}\right)  b\right]  +b\left[  \left(  \frac{\hslash\omega_{0}}{2}%
+V_{aa}\right)  a^{\ast}+V_{ba}b^{\ast}\right] \nonumber\\
&  =-ab^{\ast}\frac{\hslash\omega_{0}}{2}-ab^{\ast}V_{aa}-\left\vert
b\right\vert ^{2}V_{ab}+\left\vert a\right\vert ^{2}V_{ab}-\frac{\hslash
\omega_{0}}{2}ab^{\ast}+ab^{\ast}V_{bb}+\nonumber\\
&  -\left\vert a\right\vert ^{2}V_{ba}+a^{\ast}b\frac{\hslash\omega_{0}}%
{2}-a^{\ast}bV_{bb}+a^{\ast}b\frac{\hslash\omega_{0}}{2}+a^{\ast}%
bV_{aa}+V_{ba}\left\vert b\right\vert ^{2}\nonumber\\
&  =\left(  V_{ab}-V_{ba}\right)  \left(  \left\vert a\right\vert
^{2}-\left\vert b\right\vert ^{2}\right)  -\hslash\omega_{0}\left(  ab^{\ast
}-ba^{\ast}\right)  +\left(  V_{aa}-V_{bb}\right)  \left(  a^{\ast}b-ab^{\ast
}\right)  \text{,}%
\end{align}
that is,%
\begin{equation}
-i\hslash\frac{d\left(  ab^{\ast}+ba^{\ast}\right)  }{dt}=\left(
V_{ab}-V_{ba}\right)  \left(  \left\vert a\right\vert ^{2}-\left\vert
b\right\vert ^{2}\right)  -\hslash\omega_{0}\left(  ab^{\ast}-ba^{\ast
}\right)  +\left(  V_{aa}-V_{bb}\right)  \left(  a^{\ast}b-ab^{\ast}\right)
\text{.} \label{up2}%
\end{equation}
Therefore, using Eqs. (\ref{up}) and (\ref{up2}), it happens that the relation
$dx/dt=\Omega_{y}z-\Omega_{z}y$ holds true. The remaining two relations in Eq.
(\ref{relation}) can be verified in a similar fashion. As a final remark, we
note that $\mathbf{\dot{r}}=\mathbf{\Omega}\times\mathbf{r}$ for the
Hamiltonian \textrm{H}$(t)\overset{\text{def}}{=}(\hslash/2)\mathbf{\Omega
}\left(  t\right)  \cdot\mathbf{\boldsymbol{\sigma}}$ reduces to the condition
$\mathbf{\dot{a}=}2\mathbf{h\times a}$ for the Hamiltonian $\mathrm{H}%
(t)\overset{\text{def}}{=}\mathbf{h}(t)\cdot\mathbf{\boldsymbol{\sigma}}$ once
we consider the correspondences $\mathbf{r\leftrightarrow a}$, $\mathbf{\Omega
}/2\leftrightarrow\mathbf{h}$, and set $\hslash=1$. With this last remark, we
end our two derivations.

\end{document}